\documentclass[twocolumn,pra,superscriptaddress,showpacs,aps,floatfix]{revtex4-2}
\usepackage{color}
\usepackage{amsmath}
\usepackage{amssymb}
\usepackage{txfonts}
\usepackage{graphicx}
\usepackage{epstopdf}
\usepackage[unicode]{hyperref}
\usepackage{cancel}
\hypersetup{pdftitle={2D SOC BEC},
pdfauthor={},
pdfsubject={2D SOC BEC},
colorlinks=true,
linkcolor=red,
citecolor=magenta,
filecolor=black,
urlcolor=blue}


\begin{document}

\title{ Analysis on Leptogenesis and Neutrino Masses in the CP-Violating Standard Model}

\author{Chilong Lin }
\address{National Museum of Natural Science, 1st, Guan Chien RD., Taichung, 40453 Taiwan, ROC}

\date{Version of \today. }

\begin{abstract}

This article extends the CP-violating Standard Model (CPVSM) from the quark sector to the lepton sector to investigate leptogenesis and neutrino masses. 
Using the identity $\Delta_{hm}+\Delta_{ml} = \Delta_{hl}$,
where $\Delta_{ij} \equiv (m^2_i -m^2_j)$ denoting the mass-squared difference (MSD) between fermions $i$ and $j$, 
and $h$, $m$, and $l$ labeling the heaviest, middle, and lightest fermions of a given type, respectively, 
we predict the third neutrino MSD from two experimental inputs: $\Delta_a = 2.51 \times 10^{-3}$ eV$^2$ and $\Delta_b = 7.42 \times 10^{-5}$ eV$^2$. 
Of six possible assignments of these values to the three MSDs $\Delta_{hm}$, $\Delta_{ml}$, and $\Delta_{hl}$, four consistent cases remain and are grouped into two classes under a phenomenological point-like approach.
All four predict similar heaviest and lightest neutrino masses ($m_h \sim$ 5.01 $\times 10^{-2}$ eV and $m_l \sim$ 6.09098 $\times 10^{-3}$ eV),
but differ in the middle mass: $m_m \sim$  4.97283 $\times 10^{-2}$ eV in Class 1, and $m_m \sim$ 1.05499 $\times 10^{-2}$ eV in Class 2.
In a complementary analysis, treating the mass ratio $g\equiv m_h /m_m$ as a variable, we examine how $m_h$, $m_m$, $m_l$, and $g’\equiv m_m/m_l$ evolve with $g$.
Of particular interest are the ranges of $g$ bounded by MSD-based values derived in Subsection III-A (blue points) and values derived from a previously predicted $m_l \sim$ 8.61 $\times 10^{-3}$ eV (green points).
Finally, using the leptonic Jarlskog measure of CP violation (CPV) 
$\Delta_{CP(l)} \equiv~ J_{(l)} \cdot (\Delta_{ij} \cdot \Delta_{jk} \cdot \Delta_{ki})_{(\ell)} \cdot (\Delta_{ij} \cdot \Delta_{jk} \cdot \Delta_{ki})_{(\nu)}$, 
we find that leptogenesis is at least 71 orders of magnitude weaker than baryogenesis in the CPVSM.
This striking discrepancy suggests that new physics beyond the Standard Model (BSM) is required for leptogenesis to account for the observed Baryon Asymmetry of the Universe (BAU).

\end{abstract}

\maketitle

\section{Introduction}

There are four types of fermions in the Standard Model (SM) of electroweak interactions: up-type quarks, down-type quarks, charged leptons, and neutrinos. 
As early as 1964, physicists observed violation of CP symmetry in the decays of $K$ mesons \cite{Christenson1964}. 
More recently, theoretical studies \cite{Lin2020a, Lin2021, Lin2023} have identified general patterns in quark mass matrices that naturally give rise to  CP violation (CPV) in the SM
  through the generation of a complex phase in the Cabibbo-Kobayashi-Maskawa (CKM) matrix \cite{Cabibbo1963, KM1973}.
This type of the Standard Model is referred to as the CP-Violating Standard Model (CPVSM).   
\\

Among the four types of fermions, three generations have been identified for both quark types and charged leptons, and the masses of these nine fermions are well determined.
 In contrast, the neutrino masses remain undetermined due to their extremely small values, which make direct detection challenging. 
To date, only two mass-squared differences (MSDs) have been measured in experiments involving solar, atmospheric, reactor, and accelerator neutrinos \cite{Gonzalez2014, MACRO2001, Soudan2003, SuK1998}. These are denoted as $\Delta_a = 2.51 \times 10^{-3}$ eV$^2$ and $\Delta_b = 7.42 \times 10^{-5}$ eV$^2$ \cite{Esteban2020} in this manuscript. \\

At first glance, it appears that two measured values are insufficient to determine the three neutrino mass parameters. However, a natural constraint among the MSDs exists:
\begin{eqnarray}
\Delta_{32} + \Delta_{21} + \Delta_{13} = \Delta_{32} + \Delta_{21} - \Delta_{31} = 0,
\end{eqnarray}
where $\Delta_{ij} \equiv m_i^2 - m_j^2$ and $i \neq j$. \\

This identity implies that knowledge of any two of the three MSDs determines the third, 
provided we understand how the measured values $\Delta_a$ and $\Delta_b$ correspond to the theoretical parameters $\Delta_{ij}$. 
While the precise ordering of neutrino masses remains one of the major open questions in neutrino physics, 
all six possible mappings between experimental and theoretical MSDs can be examined to constrain the possible neutrino mass ranges. 
In this approach, two ratios $g$ and $g'$ of the neutrino masses are treated as key variables in the analysis. \\

As will be shown later, six possible correspondences exist between the experimentally measured $\Delta_a$ and $\Delta_b$ and the theoretical parameters $\Delta_{ij}$. 
 If we require all MSDs to be non-negative by definition, two of these correspondences will be excluded since the third MSD are negative in these cases. 
In this article, we use $m_h$, $m_m$, and $m_l$ to denote the heaviest, middle, and lightest neutrinos rather than the conventional $m_1$, $m_2$, and $m_3$ as the latter notation may create confusion regarding the ordering of eigenvalues. 
With this notation, the relationship mentioned in the previous paragraph can be expressed as:
\begin{eqnarray}
\Delta_{hm}+\Delta_{ml} = \Delta_{hl}, 
\end{eqnarray}
where all three MSDs are positive by definition. \\

Once the three MSDs are determined, we can predict not only the neutrino mass spectrum but also the degree of CPV in the lepton sector and the resulting leptogenesis within the SM framework. 
To quantify CPV, we adopt Jarlskog measure of CPV \cite{Jarlskog1985}, defined as:
\begin{eqnarray}
\Delta_{CP(q)} \equiv~ J_{(q)} \cdot \Delta m^2_{(u)} \cdot \Delta m^2_{(d)} \cdot T^{-12},
\end{eqnarray}
where $J_{(q)}$ is the Jarlskog invariant for the quark sector, and
\begin{eqnarray}
 \Delta m^2_{(u)} \equiv (\Delta_{hm} \cdot \Delta_{ml} \cdot \Delta_{hl})_{(u)}, ~~~~~~\Delta m^2_{(d)}  \equiv (\Delta_{hm} \cdot \Delta_{ml} \cdot \Delta_{hl})_{(d)},
\end{eqnarray}
 are the products of the three MSDs for the up-type and down-type quark sectors, respectively. 
Here, $T \sim 100$ GeV denotes the temperature of the electroweak phase transition.
In this work, we extend this formulation to the lepton sector by replacing the quark indices $(u)$, $(d)$, and $(q)$ with their lepton counterparts $(\nu)$, $(\ell)$, and $(l)$, 
corresponding to neutrinos, charged leptons, and the lepton sector as a whole. \\

By applying the same analytical procedures from the quark sector to the lepton sector, all four viable correspondences yield a consistent result:
leptogenesis is at least 71 orders of magnitude weaker than baryogenesis within the CPVSM, even when neutrino masses are allowed to be extremely large in this model.
This dramatic disparity arises primarily from the significant mass hierarchy between quarks and leptons, especially the extremely small MSDs among neutrinos. \\

This result strongly suggests that in the CPVSM, leptogenesis plays a negligible role in explaining the Baryon Asymmetry of the Universe (BAU). Consequently, if leptogenesis is to account for a significant portion of the observed BAU, new physics beyond the Standard Model (BSM) will be required.\\

However, before examining leptogenesis and neutrino masses, we briefly review the CPVSM and introduce parameter transformations related to its eigenvalues, 
which will be helpful for the subsequent analysis of leptogenesis. \\

In Section II, the CPVSM is briefly reviewed, beginning with the general fermion mass matrix pattern and culminating in the most recent analytically diagonalizable five-parameter formulation. This model naturally generates a CP-violating complex phase in the CKM matrix and reproduces the experimental CKM elements to order $O(10^{-2})$ at tree level. \\

The analysis starts with the most general 3$\times$3 mass matrix pattern containing eighteen unknown parameters.
By exploiting the fact that both $M$ and ${\bf M^2} \equiv M \cdot M^{\dagger}$ are diagonalized by the same unitary transformation $\bf U $, the problem simplifies considerably. 
Since $\bf M^2 $ is inherently Hermitian, the number of independent parameters is naturally reduced to nine. \\ 

Assuming that the real and imaginary parts of $\bf M^2$ can be diagonalized simultaneously, the number of independent parameters is further reduced to five.
This five-parameter $\bf M^2$ matrix is analytically solvable by construction, with the diagonalizing transformation matrix $\bf U$ depending on only two of the five parameters.
As a result, the CKM matrix, defined as $V_{CKM} \equiv \mathbf{U}_u \cdot \mathbf{U}_d^{\dagger}$, contains four independent parameters—two from $\bf U_u$ and two from $\bf U_d^{\dagger}$—sufficient to generate complex matrix elements.
This structure enables CPV to arise naturally and explicitly within the CPVSM.  \\

In this framework, the squared-mass eigenvalues depend on five parameters: $\bf A$, $\bf B$, $\bf C$, $\bf x$, and $\bf y$. 
These eigenvalues can be reparametrized in terms of three new variables, $\alpha$, $\beta$, and $\gamma$, 
which are composites of the original five parameters and serve as effective parameters capturing the essential degrees of freedom. 
Given the three physical squared masses for any fermion type, one can solve the resulting system of equations to determine the values of $\alpha$, $\beta$, and $\gamma$ exactly. \\

In the up-type quark, down-type quark, and charged lepton sectors, the masses of all three generations are experimentally well-established. 
Given these measured values, one can theoretically substitute them into the eigenvalue expressions to solve for the corresponding parameters \(\alpha\), \(\beta\), and \(\gamma\). 
However, since the theory does not a priori specify the correspondence between eigenvalues and fermion generations,
there exist six possible assignments of the three theoretical eigenvalues to the three observed fermion masses within each sector.
To account for this ambiguity, we systematically analyze all six possible correspondences for each fermion type and determine the resulting parameter sets. 
The computed values of \(\alpha\), \(\beta\), and \(\gamma\) for each correspondence are presented in Tables 1, 2, and 3, 
for the charged lepton, up-type quark, and down-type quark sectors, respectively. 
The analysis for the neutrino sector, which presents additional complexities, is deferred to Section III. \\

The theoretical eigenvalues exhibit mass degeneracy between two generations when $\gamma$ (proportional to $\bf C$) approaches zero.
A more pronounced degeneracy involving all three generations arises when $\beta$ also approaches zero. 
Although the time or temperature dependence of these parameters remains unknown, their potential variation allows for meaningful theoretical exploration.
Fig. 1 illustrates four possible scenarios for fermion mass evolution from the early universe to the present, noting that each fermion type may exhibit distinct evolutionary patterns. \\

According to Big Bang cosmology, the universe's temperature was extremely high in its earliest moments. 
Since higher temperatures generally correspond to greater symmetry, it is reasonable to infer that the primordial universe was so hot that all symmetries were conserved. 
As the universe expanded and cooled, these symmetries broke one after another. 
For instance, spontaneous symmetry breaking (SSB) of electroweak gauge symmetry occurred at approximately 159.5 GeV \cite{D'Onofrio2016}, generating particle masses. Under such conditions, assuming only three fermion generations, an $S_3$ symmetry among them could have existed in the very early universe. \\

Our previous CPVSM studies revealed a direct connection between CP violation and $S_N$ symmetry breaking. 
When fermion $\bf M^2$ matrices in weak interactions respect $S_3$ symmetry, CP symmetry remains conserved. 
However, when $S_3$ symmetry breaks down to $S_2$ symmetry for particular fermion types, complex phases appear in the CKM matrix. As $S_2$ symmetries further break into completely asymmetric structures, both magnitudes and phases of CKM elements vary with the parameters  $\alpha$, $\beta$, and $\gamma$. Thus, CPV emerges and evolves alongside $S_N$ symmetry breaking. \\

Notably, whether the system exhibits $S_N$ symmetry or complete asymmetry, the three mass eigenvalues can remain non-degenerate. 
Eigenvalue degeneracy depends solely on whether $\gamma = 0$, indicating that mass degeneracy is independent of $S_N$ symmetry. These topics are explored in detail in Section II. \\

The Section III is devoted to the analysis of neutrino masses and leptogenesis.
At the beginning, two neutrino mass ratios are introduced to streamline the subsequent analysis: $g \equiv m_h / m_m$ and $g’ \equiv m_m / m_l$.
Then, in Subsection III-A, all six possible correspondences between the experimentally established $\Delta_a$ and $\Delta_b$, 
and the theoretical derived--$\Delta_{hm}, ~\Delta_{ml}$, and $\Delta_{hl}$--are systematically examined. 
Since $\Delta_{hm}, ~\Delta_{ml}$, and $\Delta_{hl}$ are defined to be non-negative, any assignment resulting in a negative MSD is inherently inconsistent.
Consequently, four of the six configurations are found to be logically self-consistent, while the remaining two ($\bf Case~3~and~6$) are excluded. \\

The analysis then focuses on the four viable cases.
$\bf Case~1~and ~ 5$, grouped as Class 1, both yield the hierarchy $m^2_h \sim m^2_m \gg m^2_l$, indicating that $m^2_m$ is close to $m_h^2$ and well separated from $m_l^2$. 
In contrast, $\bf Case~2~and ~ 4$, designated as Class 2, produce the hierarchy $m^2_h \gg m^2_m \sim m^2_l$, showing that $m^2_m$ is close to $m_l^2$ and distant from $m_h^2$. 
Following this method, we categorize the four viable cases into two classes. \\

In Class 1 ($\bf Case~1~and ~5$), $m^2_m \sim m^2_h$, while $m^2_l$ is much smaller than both and is therefore assumed to be negligible.
As a result, $\Delta_a$ is close in value to both $m^2_m$ and $m^2_h$, and may lie between them.  
Since the exact position of $\Delta_a$ is unknown, 
we adopt the midpoint $\Delta_a= (m^2_h+ m^2_m)/ 2$ as a reference to compute the individual neutrino masses and the parameters $g$ and $g’$.
The results obtained using this approximation are expected to remain close to the true values. 
 Both cases yield $m_h$=5.04688 $\times 10^{-2}$ eV and $m_m$= 4.97283 $\times 10^{-2}$ eV.
However, $\bf Case~ 5$ results in $m_l$= 6.09098$i \times 10^{-3}$ eV, an unphysical imaginary value,
 indicating that the underlying assumption is not valid in $\bf Case~5$. \\

For Class 2 ($\bf Case~2~and~ 4$), $m^2_m \sim m^2_l$, and both are much smaller than $m^2_h$.
Therefore, we take the midpoint $ \Delta_b =(m^2_l+ m^2_m)/ 2 $ as the reference point. 
Upon substitution, the two cases differ only slightly in the value of $m_h$: 5.04688 $\times 10^{-2}$ eV in $\bf Case~2$ and 5.11968 $\times 10^{-2}$ eV in $\bf Case~4$. 
In contrast, both cases yield identical values of $m_m =$ 1.05499 $\times 10^{-2}$ eV and $m_l =$ 6.09098 $\times 10^{-3}$ eV. \\

In scenarios with limited experimental input, a common phenomenological approach involves selecting representative values within the theoretically allowed parameter space to facilitate further analysis. For instance, in Class 1, we impose $(\Delta_h + \Delta_m)/2 = \Delta_a$, while in Class 2, the condition $(\Delta_m + \Delta_l)/2 = \Delta_b$ is adopted. Under these respective assumptions, the predicted values for the intermediate neutrino mass $m_m$ are 4.97283 $\times 10^{-2}$ eV in Class 1 and 1.05499 $\times 10^{-2}$ eV in Class 2. \\

Remarkably, three of the four phenomenologically viable configurations--$\bf Cases ~1,~ 2,~ and~4$--yield a consistent prediction for the lightest neutrino mass, $m_l = 6.09098 \times 10^{-3}$ eV.
In contrast, $\bf Case~5$ results in a purely imaginary $m_l$, with the same absolute value. 
This predicted value of $m_l$ is somewhat lower than the global fit estimate  $m_1 = 8.61 \times 10^{-3}$ eV, as reported in \cite{Esteban2020}--
a discrepancy that remains to be tested by future experimental data.  \\

This point-wise trial-and-error method is inherently limited and may overlook viable solutions;
 it is thus best regarded as a preliminary approach in the absence of stronger constraints. 
In contrast, Subsection III-C adopts a more systematic analysis by treating $g$ as a continuous variable, 
thereby providing broader coverage and a more complete picture than the discrete method used in III-A. \\

In Subsection III-B, using the determined neutrino mass parameters, we analyze the mass hierarchies across the four fermion types and compute the twelve corresponding MSDs. 
The resulting neutrino mass ratios are found to be significantly smaller than those in the other three fermion sectors. 
Substituting the twelve MSDs into the Jarlskog measure of CPV (as defined in Eq.(3)), 
we find that the sum of the two MSD products in the quark sector exceeds that of the lepton sector by approximately 74 orders of magnitude. 
This suggests that leptogenesis driven by the Dirac CP-violating phase is negligible in the present universe compared to baryogenesis from the quark sector. 
Even after accounting for the Jarlskog invariant, the disparity remains larger than 71 orders of magnitude. \\ 

In Subsection III-C, neutrino masses are examined from an alternative perspective, yielding results that closely corroborate those presented in Subsection III-A.
For each scenario considered, analytical expressions are derived to illustrate how $m_h$, $m_m$, $m_l$, and  $g'$ depend on the scaling factor $g$. \\

Among the four cases studied, in $\bf Case~1~and ~5$, $g'$ increases from 1 and diverges to infinity as $g$ approaches critical values of 1.014512 and 1.01467, respectively. 
Similarly, in $\bf Case~2~ and~ 4$, $g'$ also increases from 1 and diverges as $g$ approaches 5.81614 and 5.90148, respectively. 
Beyond these critical thresholds, the emergence of negative or imaginary mass values imposes physical constraints, limiting the viable parameter space. \\

Consequently, the neutrino masses are constrained to lie within narrow, well-defined ranges of $g$, offering potential guidance for the design of future experimental searches.
The detailed behaviors of  $m_h$, $m_m$, $m_l$, and  $g'$ as functions of $g$ are plotted and discussed in Subsection III-C. \\

Section IV is dedicated to conclusions and discussions. \\

\section{CPVSM and Mass Degeneracy Phenomena}\label{sect2}

In this section, the CP-violating Standard Model (CPVSM) is briefly reviewed, along with some supplementary insights.
The model starts from the most general pattern of the fermion mass matrices, $M$, 
and a mathematical relation between $M$ and its square, ${\bf M^2} \equiv M \cdot M^{\dagger}$, showing that both are diagonalized by the same unitary matrix, $\bf U$.
Since $\bf M^2$  is naturally Hermitian, the complexity of the problem is significantly reduced, as $\bf M^2$  involves only nine parameters, compared to the eighteen parameters in $M$. \\

By assuming that the real and imaginary parts of $\bf M^2$ can be diagonalized simultaneously by $\bf U$, the number of independent parameters is further reduced from nine to five.
At this stage, the $\bf M^2$ matrix becomes analytically diagonalizable, and a complex phase naturally emerges in the resulting CKM matrix  \cite{Lin2021, Lin2023}. 
This provides a special solution to the problem of CPV origin in the Standard Model, though it is not yet fully complete.
Therefore,  it is logical to extend this approach to the lepton sector, in order to investigate whether a similar mechanism could also lead to CP violation in that context. 
Even further, to see how leptogenesis contribute to the production of Baryon Asymmetry of the Universe? \\

As shown in \cite{Lin2019, Lin2020a, Lin2021, Lin2023}, the most general 3$\times$3 mass matrix pattern can always be given by
\begin{eqnarray}
M &=& \left( \begin{array}{ccc}A_1+i D_1 &B_1+i C_1 & B_2+ i C_2 \\ B_4 +i C_4 & A_2 +i D_2 & B_3 +i C_3 \\ B_5 +i C_5 & B_6 + i C_6 & A_3 +i D_3 \end{array}\right) \nonumber \\
&=& M_R + i ~M_I =\left( \begin{array}{ccc}A_1 &B_1 & B_2 \\ B_4 & A_2 & B_3 \\ B_5 & B_6 & A_3 \end{array}\right)+ i~ \left( \begin{array}{ccc} D_1 & C_1 & C_2 \\ C_4 & D_2 & C_3 \\ C_5 & C_6 & D_3 \end{array}\right),
\end{eqnarray}
in which there are in total eighteen independent parameters, nine from the real coefficients and nine from the imaginary coefficients of its nine elements. 
Such a pattern is obviously too complicated to be diagonalized analytically. \\

However, the eigenvectors or the unitary matrix that diagonalizes the $M$ matrix are the same as those of the mass-squared matrix ${\bf M^2}\equiv M \cdot M^{\dagger}$.
The general pattern of $\bf M^2$ is given by 
\begin{eqnarray}
\bf M^2 &=& \left( \begin{array}{ccc}{\bf A_1} & {\bf B_1+i C_1} & {\bf B_2+ i C_2} \\ {\bf B_1 -i C_1} & {\bf A_2} & {\bf B_3 +i C_3} \\ {\bf B_2 -i C_2} & {\bf B_3 - i C_3} & {\bf A_3} \end{array}\right) \nonumber \\
&=& {\bf M^2_R}+i~ {\bf M^2_I} = \left( \begin{array}{ccc}{\bf A_1} & {\bf B_1} & {\bf B_2} \\ {\bf B_1} & {\bf A_2} & {\bf B_3} \\ {\bf B_2} & {\bf B_3} & {\bf A_3} \end{array}\right) 
+ i ~ \left( \begin{array}{ccc}0 & {\bf C_1} & {\bf C_2} \\ {\bf -C_1} & 0 & {\bf C_3} \\ {\bf - C_2} & {\bf -C_3} & 0 \end{array}\right),
\end{eqnarray}
where the boldface parameters $\bf A$, $\bf B$, and $\bf C$ are composed of the parameters in $M$ as follows:
\begin{eqnarray}
{\bf A_1} &=& A_1^2 + D_1^2 + B_1^2 + C_1^2 + B_2^2 + C_2^2, \\
{\bf A_2} &=& A_2^2 + D_2^2 + B_3^2 + C_3^2 + B_4^2 + C_4^2, \\
{\bf A_3} &=& A_3^2 + D_3^2 + B_5^2 + C_5^2 + B_6^2 + C_6^2, \\
{\bf B_1} &=& A_1 B_4 + D_1 C_4 + B_1 A_2 + C_1 D_2 + B_2 B_3 +C_2 C_3, \\
{\bf B_2} &=& A_1 B_5 + D_1 C_5 + B_1 B_6 + C_1 C_6 + B_2 A_3 +C_2 D_3, \\
{\bf B_3} &=& B_4 B_5 + C_4 C_5 + B_6 A_2 + C_6 D_2 + A_3 B_3 +D_3 C_3, \\
{\bf C_1} &=& D_1 B_4 -A_1 C_4 +A_2 C_1 -B_1 D_2 +B_3 C_2 -B_2 C_3, \\
{\bf C_2} &=& D_1 B_5 -A_1 C_5 +B_6 C_1 -B_1 C_6 +A_3 C_2 -B_2 D_3, \\
{\bf C_3} &=& C_4 B_5 -B_4 C_5 +D_2 B_6 -A_2 C_6 +A_3 C_3 -B_3 D_3.
\end{eqnarray}
Thus, only nine real parameters remain independent since $\bf M^2$ is naturally Hermitian, regardless of whether $M$ is Hermitian or not. \\

Obviously, diagonalizing the nine-parameter $\bf M^2$ matrix analytically remains impractical.
However, as demonstrated in \cite{Lin2021}, assuming that both $\bf M^2_R$ and $\bf M^2_I$ can be diagonalized  simultaneously by the same unitary matrix $\bf U$,
 four extra constraints arise among the parameters. 
This reduces the number of independent parameters from nine to five. \\

While this method leads to an analytic solution, it is not the most general one, as imposing additional assumptions or constraints reduces the solution's generality. 
Nonetheless, the assumption used here represents the weakest constraint achievable with current techniques. \\

Defining  $\bf A \equiv A_3$, $\bf B \equiv B_3$, $\bf C \equiv C_3$, $\bf x \equiv {\bf B_2 \over B_3}$, and $\bf y \equiv{\bf B_1 \over B_3}$ as the five remaining free parameters and replacing all others accordingly, 
the eigenvalues are given by:
\begin{eqnarray}
{\bf m^2_1} &=& ({\bf A- {x \over y}B) - {\sqrt{\bf x^2 +y^2 +x^2 y^2} \over {x y}} C}, \\
{\bf m^2_2} &=& ({\bf A-{x \over y} B) + {\sqrt{\bf x^2 +y^2 +x^2 y^2} \over {x y}}C}, \\
{\bf m^2_3} &=& {\bf A+{{(x^2+1) y} \over x} B} \nonumber \\
&=& ({\bf A- {x \over y}B) + {{x^2 y^2 + x^2 +y^2}\over {x y }} B},
\end{eqnarray}
while the $\bf U$ matrix is given by:
\begin{eqnarray}
{\bf U} = ~\left( \begin{array}{ccc}
{-\sqrt{\bf x^2+y^2} \over \sqrt{\bf 2(x^2+y^2+x^2 y^2)}} & {\bf {x(y^2-i \sqrt{\bf x^2+y^2+x^2 y^2})} \over {\bf \sqrt{2} \sqrt{\bf x^2+y^2} \sqrt{\bf x^2+y^2+x^2 y^2}}} & {\bf {y(x^2+i \sqrt{\bf x^2+y^2+x^2 y^2})} \over {\bf \sqrt{\bf 2} \sqrt{\bf x^2+y^2} \sqrt{\bf x^2+y^2+x^2 y^2}}} \\
{-\sqrt{\bf x^2+y^2} \over \sqrt{\bf 2(x^2+y^2+x^2 y^2)}} & {{\bf x(y^2+i \sqrt{\bf x^2+y^2+x^2 y^2})} \over {\sqrt{\bf 2} \sqrt{\bf x^2+y^2} \sqrt{\bf x^2+y^2+x^2 y^2}}} &~{{\bf y(x^2-i \sqrt{\bf x^2+y^2+x^2 y^2})} \over {\sqrt{\bf 2} \sqrt{\bf x^2+y^2} \sqrt{\bf x^2+y^2+x^2 y^2}}} \\
{{\bf x y} \over \sqrt{\bf x^2+y^2+x^2 y^2}} &~ {\bf y \over \sqrt{\bf x^2+y^2+x^2 y^2}} &~{\bf x \over \sqrt{\bf x^2+y^2+x^2 y^2}} \end{array}\right).
\end{eqnarray}

It is noteworthy that in such a model, the $\bf U$ matrix depends on only two of the five remaining parameters.
Additionally, it is important to emphasize that the mass-squared eigenvalues $\bf m^2_1$, $\bf m^2_2$, and $\bf m^2_3$ may correspond to the physical fermion masses in various ways.
As previously noted, the heaviest, intermediate, and lightest fermions of a given type are denoted by $m_h$, $m_m$, and $m_l$, respectively. 
The various permutations by which the three eigenvalues may be associated with the three physical masses will be systematically investigated.   \\

In such a parameterization, the $\bf M^2$ matrix can be further expressed as
\begin{eqnarray}
{\bf M^2} &=&
\left( \begin{array}{ccc} {\bf A +  (x y- {x \over y})B} & {\bf y B} & {\bf x B} \\ {\bf y B} & {\bf A + ({y \over x}-{x \over y})B} & {\bf B} \\ {\bf x B} & {\bf B} & {\bf A} \end{array}\right) \nonumber \\
&+& i~\left( \begin{array}{ccc} 0 & {\bf  {1 \over y}C} & - {\bf  {1 \over x}C} \\ - {\bf  {1 \over y}C} & 0 & {\bf C} \\ {\bf {1 \over x}C} & - {\bf C} & 0 \end{array}\right) .
\end{eqnarray}

In Eq.(16)-(18), there are five parameters in three eigenvalues but only three given masses in each fermion type. 
Thus, it is clearly impossible to determine the details of any of the parameters conclusively. 
However, if we denote the $\bf A$- and $\bf B$-relative parts in the forefront brackets of Eq.(16)-(18) as $\alpha$, 
the latter $\bf C$-relative parts as $\gamma$, and $\beta \equiv {\bf m^2_3} -\alpha$,
the eigenvalues in Eq.(16)-(18) can be thus revised as follows:
\begin{eqnarray}
{\bf m^2_1} =~ \alpha -\gamma,~~~~~{\bf m^2_2}&= &~ \alpha +\gamma, ~~~~~{\bf m^2_3} =~ \alpha + \beta, \\
{\bf m^2_1 + m^2_2 + m^2_3} &=& 3\alpha+\beta
\end{eqnarray}
where
\begin{eqnarray}
\alpha &=&~{\bf (m^2_1+m^2_2) \over 2}= {\bf A- {x \over y}B}, \\
\beta &=&~{\bf m^2_3 - {(m^2_1+m^2_2) \over 2}}= {\bf {(x^2 y^2 + x^2 +y^2) \over {x y }} B} , \\
\gamma &=&~ {\bf (m^2_2 -m^2_1) \over 2}}= {\bf {\sqrt{\bf x^2 +y^2 +x^2 y^2} \over {x y}} C .
\end{eqnarray}

In this manner, the parameters $\alpha$, $\beta$, and $\gamma$ can be determined by the experimentally given fermion masses,
 to built a direct connection between the theoretical eigenvalues and the physical fermion masses.
This approach obviously applies to the quark sector and charged leptons and potentially to neutrinos as well.
Such a simplification will be helpful in the coming analyses to be shown below. \\

In Eq.(22), it is evident that the sum of the three mass-squares depends only on the parameters  $\alpha$ and $\beta$. 
The variation of $\gamma$, if it does vary, does not affect the sum of the three mass-squares for a given fermion type. 
Interestingly, two of the eigenvalues become degenerate when ${\bf C}$=0 (which makes $\gamma=0$), with splitting occurring only when $\gamma$ becomes non-trivial. \\

It is evident from Eq.(25) that the first term of $\gamma$ satisfies ${\sqrt{x^2 +y^2+x^2 y^2}\over {x~y}} > 1$ for arbitrary values of $x$ and $y$.
This inequality implies that $\gamma$ vanishes only when $\bf C=0$.
Therefore, if $\bf C$ is non-zero, degeneracy does not occur--that is, the eigenvalues are split.
However, the underlying mechanism responsible for generating a non-trivial $\bf C$ remains unclear.
It is plausible that this mechanism is related to the temperature of the universe, 
as many physical phenomena exhibit symmetry breaking below certain critical temperature thresholds. \\

Regardless how the eigenvalues lose their degeneracy, there are at most four possible relationships among them, as illustrated in Fig. 1,
which shows  the evolution of the eigenvalues with temperature.
These configurations can be categorized into two distinct groups, which represent limiting cases that occur when $\gamma$=0—that is, before full mass splitting:\\

Group 1: $m_2=m_1 > m_3$ (Fig. 1-1 and Fig. 1-3) \\

Group 2: $m_3 > m_2=m_1$ (Fig. 1-2 and Fig. 1-4) \\

Within each group, there are two distinct ways in which the initially degenerate states can further split: \\

(i) In one scenario, $m_1$ and $m_2$ split, and one of them evolves to surpass $m_3$; this can occur whether the initially degenerate pair lies above (Fig. 1-3) or below (Fig. 1-4) $m_3$. \\

(ii) In the other scenario, $m_1$ and $m_2$ split but remain on the same side of $m_3$, never intersecting its trajectory (Fig. 1-1 and Fig. 1-2). \\

\begin{figure}

\includegraphics{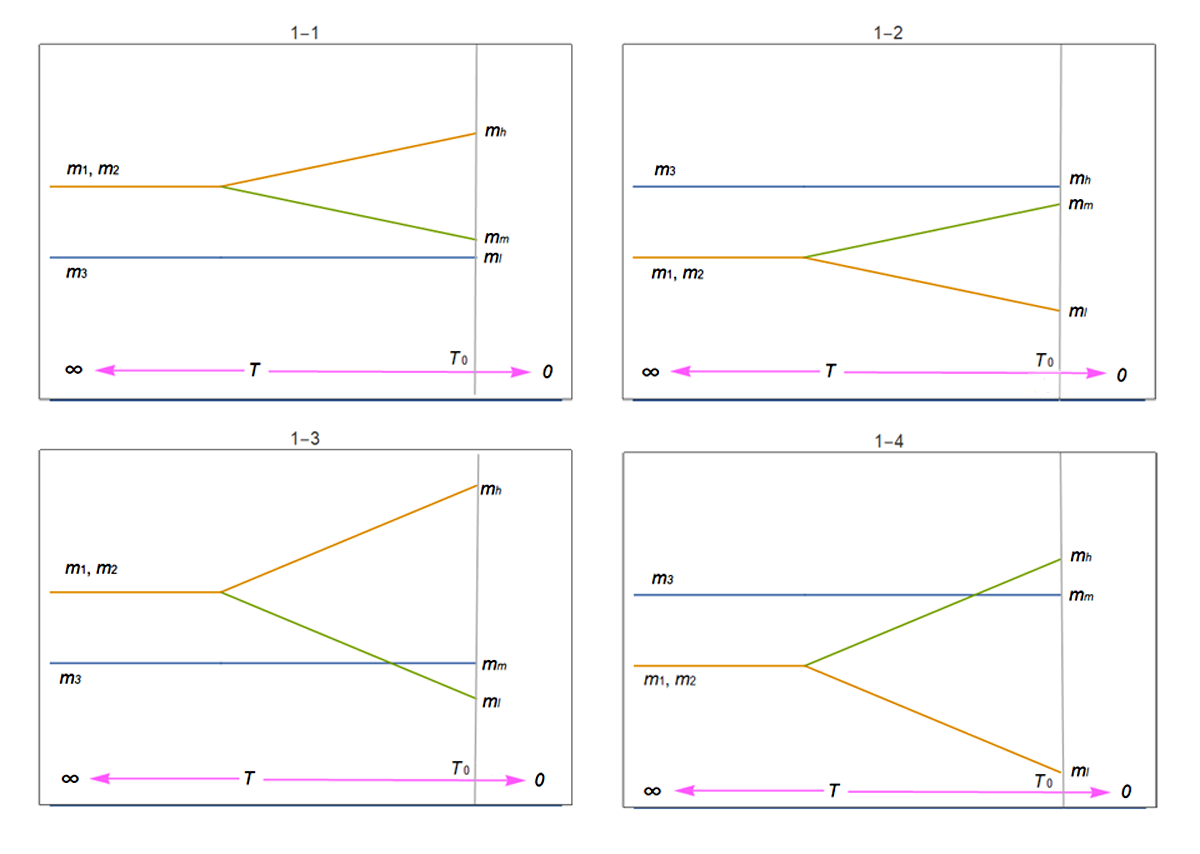}\label{fig:petit3leaks3D}
\caption{ Four ways the two degenerate eigenvalues split as $\gamma$ grows from zero. 
The horizontal axis is the temperature $T$.
 $T_0$ is the present temperature; at the beginning of the universe, $T \rightarrow \infty $, and it will approach zero as the universe expands. 
These figures can be divided into two groups: one in which $m_2=m_1 > m_3$ (Fig. 1-1 and Fig. 1-3) and the other in which $m_3 > m_2=m_1 $ (Fig. 1-2 and Fig. 1-4) when $\gamma=0$.
In each group, there are two possible ways the degenerate states can split: 
either one of the originally degenerate states grows to surpass $m_3$, which could be originally lower (Fig. 1-3) or higher (Fig. 1-4) than the degenerate $m_1=m_2$ state, 
or $m_1$ and $m_2$ never surpass the line of $m_3$ (Fig. 1-1 and Fig. 1-2).
It's important to note that $m_3$ may not always be a fixed value as shown in the figures since $\beta$ has no reason to remain invariant. 
However, the $m_3$ lines in the figures are simply sketches to illustrate the relationship between $m_3$ and the other two masses. 
}
\label{figure1}

\end{figure}

In this way, the parameters $\alpha$, $\beta$, and $\gamma$ can be determined unambiguously by substituting experimentally measured fermion masses into Eqs. (23)–(25).
For example, when applied to the charged lepton sector, the input values are $m_l=m_e$ = 0.000511 GeV, $m_m =m_{\mu}$ = 0.1057 GeV, and $m_h=m_{\tau}$ = 1.7768 GeV. 
However, there are six distinct ways to assign the eigenvalues $m^2_1$, $m^2_2$, and $m^2_3$ to the physical squared masses $m^2_l$, $m^2_m$, and $m^2_h$, as listed below:
\begin{eqnarray}
Case~A.~ (m^2_1,~m^2_2,~m^2_3)~&\rightarrow& ~(m^2_l,~m^2_m,~m^2_h), \\
Case~B. ~(m^2_1,~m^2_2,~m^2_3)~&\rightarrow& ~(m^2_l,~m^2_h,~m^2_m), \\
Case~C. ~(m^2_1,~m^2_2,~m^2_3)~&\rightarrow& ~(m^2_m,~m^2_l,~m^2_h), \\
Case~D. ~(m^2_1,~m^2_2,~m^2_3)~&\rightarrow& ~(m^2_m,~m^2_h,~m^2_l), \\
Case~E. ~(m^2_1,~m^2_2,~m^2_3)~&\rightarrow& ~(m^2_h,~m^2_m,~m^2_l), \\
Case~F. ~(m^2_1,~m^2_2,~m^2_3)~&\rightarrow& ~(m^2_h,~m^2_l,~m^2_m), 
\end{eqnarray}
Each of these cases will be discussed in detail in the following analysis. \\

\subsection{$ (m^2_1,~m^2_2,~m^2_3)~\rightarrow ~(m^2_l,~m^2_m,~m^2_h)$}\label{sect2-A}

Taking the charged leptons for an example, $m_l=m_e$ = 0.000511 GeV, $m_m =m_{\mu}$=0.1057 GeV, and $m_h=m_{\tau}$= 1.7768 MeV.
In this case 
\begin{eqnarray}
\alpha_{\ell 1} ~&=&~ {{m^2_{\mu} +m^2_e } \over 2} ~=~5.58638 ~\times ~10^{-3}~{\rm GeV}^2, \\
\beta_{\ell 1} ~&=&~ m^2_{\tau} -\alpha_{\ell 1} ~=~ {{ ({\bf x^2}_{\ell} {\bf y^2}_{\ell} + {\bf x^2}_{\ell} + {\bf y^2}_{\ell} ) } \over { {\bf x}_{\ell} {\bf y}_{\ell}}} {\bf B}_{\ell} ~\nonumber \\ 
&=&~3.15214~{\rm GeV}^2, \\
\gamma_{\ell 1} ~&=&~ {{m^2_{\mu} -m^2_e } \over 2} =~{ \sqrt{{\bf x^2}_{\ell} + {\bf y^2}_{\ell} + {\bf x^2}_{\ell} {\bf y^2}_{\ell} } \over {{\bf x}_{\ell} {\bf y}_{\ell}}} {\bf C}_{\ell} ~\nonumber \\ 
&=&~5.58611 ~\times~10^{-3}~{\rm GeV}^2,
\end{eqnarray}
where the subindex ${\ell}$ stands for the charge leptons. 
However, the parameters $\bf A_{\ell}$, $\bf B_{\ell}$, $\bf C_{\ell}$, $\bf x_{\ell}$, and $\bf y_{\ell}$ are not determined. \\

\subsection{$(m^2_1,~m^2_2,~m^2_3)~\rightarrow ~(m^2_l,~m^2_h,~m^2_m)$}\label{sect2-B}

\begin{eqnarray}
\alpha_{\ell 2} ~&=&~{m^2_{\tau}+{m^2_e }\over 2} ~=~1.57886~{\rm GeV}^2, \\
\beta_{\ell 2} ~&=&~ m^2_{\mu} -\alpha_{\ell 2} ~=~-1.56769~{\rm GeV}^2, \\
\gamma_{\ell 2} ~&=&~ {{m^2_{\tau} -m^2_e } \over 2} ~=~1.57886~ {\rm GeV}^2.
\end{eqnarray}

\subsection{$ (m^2_1,~m^2_2,~m^2_3)~\rightarrow ~(m^2_m,~m^2_l,~m^2_h)$}\label{sect2-C}

\begin{eqnarray}
\alpha_{\ell 3} ~&=&~{{m^2_e + m^2_{\mu}}\over 2} ~=~5.58638 ~\times~10^{-3}~{\rm GeV}^2, \\
\beta_{\ell 3} ~&=&~ m^2_{\tau} -\alpha_{\ell 3}~=~3.15214~{\rm GeV}^2, \\
\gamma_{\ell 3} ~&=&~ {{m^2_e-m^2_{\mu} } \over 2} ~=~-5.58611 ~\times~10^{-3}~{\rm GeV}^2.
\end{eqnarray}

\subsection{$(m^2_1,~m^2_2,~m^2_3)~\rightarrow ~(m^2_m,~m^2_h,~m^2_l)$}\label{sect2-D}

\begin{eqnarray}
\alpha_{\ell 4} ~&=&~{{m^2_{\tau} + m^2_{\mu}}\over 2} ~=~1.58445~ {\rm GeV}^2, \\
\beta_{\ell 4} ~&=&~ m^2_e -\alpha_{\ell 4}~=~-1.58445~{\rm GeV}^2, \\
\gamma_{\ell 4} ~&=&~ {{m^2_{\tau}-m^2_{\mu} } \over 2} ~=~1.57328~{\rm GeV}^2.
\end{eqnarray}

\subsection{$ (m^2_1,~m^2_2,~m^2_3)~\rightarrow ~(m^2_h,~m^2_m,~m^2_l)$}\label{sect2-E}

\begin{eqnarray}
\alpha_{\ell 5} ~&=&~{{m^2_{\mu} +m^2_{\tau}}\over 2} ~=~1.58445~ {\rm GeV}^2, \\
\beta_{\ell 5} ~&=&~ m^2_e -\alpha_{\ell 5}~=~-1.58445~ {\rm GeV}^2, \\
\gamma_{\ell 5} ~&=&~ {m^2_{\mu} -{m^2_{\tau} } \over 2} ~=~-1.57328~ {\rm GeV}^2.
\end{eqnarray}

\subsection{$ (m^2_1,~m^2_2,~m^2_3)~\rightarrow ~(m^2_h,~m^2_l,~m^2_m)$}\label{sect2-F}

\begin{eqnarray}
\alpha_{\ell 6} ~&=&~{{m^2_e +m^2_{\tau}}\over 2} ~=~1.57886~{\rm GeV}^2, \\
\beta_{\ell 6} ~&=&~ m^2_{\mu} -\alpha_{\ell 6}~=~-1.56769~ {\rm GeV}^2, \\
\gamma_{\ell 6} ~&=&~ {m^2_e -{m^2_{\tau} } \over 2} ~=~-1.57886~{\rm GeV}^2.
\end{eqnarray}
\\

Similarly, the $\alpha$, $\beta$, and $\gamma$ parameters for other fermion types can be determined in the same manner. 
For instance, these parameters for the charged leptons, up-type quarks, and down-type quarks are provided in Tables I, II, and III, respectively.
In the next section, we will also examine these parameters in the neutrino sector.\\ 

\begin{table}[tbp]
\centering
\begin{tabular}{l|llllll}
$ (m^2_1,~m^2_2,~m^2_3) \setminus param.$ & ~~~~~~~~$\alpha_{\ell}$ & ~~~~~~~~$\beta_{\ell}$ & ~~~~~~~~$\gamma_{\ell}$ \\ \hline
{\tt $ \begin{array}{cccccc} (m^2_e,~m^2_{\mu},~m^2_{\tau}) \\ (m^2_e,~m^2_{\tau},~m^2_{\mu}) \\ (m^2_{\mu},~m^2_e,~m^2_{\tau}) \\ (m^2_{\mu},~m^2_{\tau},~m^2_e) \\ (m^2_{\tau},~m^2_{\mu},~m^2_e) \\ (m^2_{\tau},~m^2_e,~m^2_{\mu}) \end{array} $} &
{\tt $ \begin{array}{cccccc} 5.58638 \times 10^{-3} \\ 1.57886 \\ 5.58638 \times 10^{-3} \\ 1.58445 \\ 1.58445 \\ 1.57886 \end{array} $} &
$ \begin{array}{cccccc} 3.15214 \\ -1.56769 \\ 3.15214 \\ -1.58445 \\ -1.58445 \\ -1.56769 \end{array}$ &
$ \begin{array}{cccccc} 5.58611 \times 10^{-3} \\ 1.57886 \\ -5.58611 \times 10^{-3} \\ 1.57328 \\ -1.57328 \\ -1.57886 \end{array} $ \\ \hline
\end{tabular}

\caption{\label{tab:i} 
Parameters in the charged lepton sector. 
The masses employed here are $m_e$= 0.000511 GeV, $m_{\mu}$=0.10566 GeV, and $m_{\tau}$= 1.7768 GeV.}
\end{table}

\begin{table}[tbp]
\centering
\begin{tabular}{l|llllll}
$ (m^2_1,~m^2_2,~m^2_3) \setminus param.$ & ~~~~~~~~$\alpha_u$ & ~~~~~~~~$\beta_u$ & ~~~~~~~~$\gamma_u$ \\ \hline
{\tt $ \begin{array}{cccccc} (m^2_u,~m^2_c,~m^2_t) \\ (m^2_u,~m^2_t,~m^2_c) \\ (m^2_c,~m^2_u,~m^2_t) \\ (m^2_c,~m^2_t,~m^2_u) \\ (m^2_t,~m^2_c,~m^2_u) \\ (m^2_t,~m^2_u,~m^2_c) \end{array} $} &
{\tt $ \begin{array}{cccccc} 0.812815 \\ 15000.9 \\ 0.812815 \\ 15001.7 \\ 15001.7 \\ 15000.9 \end{array} $} &
$ \begin{array}{cccccc} 30000.9 \\ -14999.2 \\ 30000.9 \\ -15001.7 \\ -15001.7 \\ -15000.0 \end{array}$ &
$ \begin{array}{cccccc} 0.81281 \\ 15000.9 \\ -0.81281 \\ 15000.0 \\ -15000.0 \\ -15000.9 \end{array} $ \\ \hline
\end{tabular}

\caption{\label{tab:i} 
Parameters in the up-type quark sector. 
The masses employed here are $m_u$= 0.0023 GeV, $m_c$=1.275 GeV, and $m_t$= 173.21 GeV.}
\end{table}

\begin{table}[tbp]
\centering
\begin{tabular}{l|llllll}
$ (m^2_1,~m^2_2,~m^2_3) \setminus param.$ & ~~~~~~~~$\alpha_d$ & ~~~~~~~~$\beta_d$ & ~~~~~~~~$\gamma_d$ \\ \hline
{\tt $ \begin{array}{cccccc} (m^2_d,~m^2_s,~m^2_b) \\ (m^2_d,~m^2_b,~m^2_s) \\ (m^2_s,~m^2_d,~m^2_b) \\ (m^2_s,~m^2_b,~m^2_d) \\ (m^2_b,~m^2_s,~m^2_d) \\ (m^2_b,~m^2_d,~m^2_s) \end{array} $} &
{\tt $ \begin{array}{cccccc} 4.52402 \times 10^{-3} \\ 8.73621 \\ 4.52402 \times 10^{-3} \\ 8.74071 \\ 8.74071 \\ 8.73621 \end{array} $} &
$ \begin{array}{cccccc} 17.4679 \\ -8.72719 \\ 17.4679 \\ -8.74069 \\ -8.74069 \\ -8.73169 \end{array}$ &
$ \begin{array}{cccccc} 4.50098 \times 10^{-3} \\ 8.73619 \\ -4.50098 \times 10^{-3} \\ 8.73169 \\ -8.73169 \\ -8.73169 \end{array} $ \\ \hline
\end{tabular}

\caption{\label{tab:i} 
Parameters in the down-type quark sector. 
The masses employed here are $m_d$= 0.0048 GeV, $m_s$=0.095 GeV, and $m_b$= 4.180 GeV.}
\end{table}

Although the parameters $\mathbf{A}$, $\mathbf{B}$, $\mathbf{C}$, $\mathbf{x}$, and $\mathbf{y}$ are not conclusively determined by the fermion masses at this stage,
 the parameters $\alpha$, $\beta$, and $\gamma$ are fixed. 
These parameters exhibit a degeneracy between two of the three eigenvalues when $\mathbf{C} = 0$,
 and an even more pronounced degeneracy involving all three generations when $\beta =\gamma = 0$,
 resulting in $m_1 = m_2 = m_3 = \alpha$. 
The evolution of the eigenvalues with respect to $\gamma$ and $\beta$ over time or temperature could be of significant interest and warrants further investigation. \\

%
%
\section{Neutrino Masses and Leptogenesis}\label{sect3} 

Among the four types of fermions in the Standard Model, the masses of three—the two quark sectors and the charged leptons—are well determined. 
However, for neutrinos, only two MSDs are currently known, making it difficult to determine their absolute mass values. 
This section presents a more detailed investigation of the neutrino sector, with the aim of constraining the possible mass range of the three neutrino mass eigenstates.\\

Prior to advancing the analysis, we introduce a set of notational conventions to enhance clarity in the subsequent derivations.
As previously noted, the heaviest, intermediate, and lightest fermions of a given type are denoted by  $m_h$, $m_m$, and $m_l$,  respectively. 
In addition, we define two mass ratios:
\begin{eqnarray}
m_h &=& g \cdot m_m, \nonumber \\
m_m &=& g' \cdot m_l,
\end{eqnarray}
which will prove useful in the forthcoming calculations. \\

\subsection{Analysis of Neutrino Mass-Squared Differences}\label{sect3A}

The two experimentally obtained MSDs are denoted as 
\begin{eqnarray}
\Delta_a =2.51 \times 10^{-3}~ {\rm eV}^2~~{\rm and}~~~~~\Delta_b=7.42 \cdot~ 10^{-5}~ {\rm eV}^2.
\end{eqnarray}
The three theoretical defined MSDs are expressed as $\Delta_{hm} = (m^2_h - m^2_m) \ge 0$, $\Delta_{hl} = (m^2_h - m^2_l) \ge 0$, and $\Delta_{ml} = (m^2_m - m^2_l) \ge 0$,
 all of which are non-negative by definition. 
These quantities satisfy the relation $\Delta_{hm} + \Delta_{ml} = \Delta_{hl}$, as stated in Eq.(2). \\

There are six possible correspondences between the experimental MSDs, $\Delta_a$ and $\Delta_b$, 
and the three theoretical MSDs, $\Delta_{hl}$, $\Delta_{ml}$, and $\Delta_{hm}$, as outlined below: \\

$\bf Case~ 1$: $\Delta_{hl} \Leftrightarrow \Delta_a$ and $\Delta_{hm} \Leftrightarrow \Delta_b$. Then $\Delta_{ml}=\Delta_a -\Delta_b > 0$. \\

$\bf Case~ 2$: $\Delta_{hl} \Leftrightarrow \Delta_a$ and $\Delta_{ml} \Leftrightarrow \Delta_b$. Then $\Delta_{hm}=\Delta_a -\Delta_b > 0$. \\

$\bf Case~ 3$: $\Delta_{hm} \Leftrightarrow \Delta_a$ and $\Delta_{hl} \Leftrightarrow \Delta_b$. This case is excludedsince it would imply $\Delta_{hm}=\Delta_b -\Delta_a < 0$, contradicting the definition given above. \\

$\bf Case~ 4$: $\Delta_{hm} \Leftrightarrow \Delta_a$ and $\Delta_{ml} \Leftrightarrow \Delta_b$. Then $\Delta_{hl}=\Delta_a +\Delta_b > 0$. \\

$\bf Case~ 5$: $\Delta_{ml} \Leftrightarrow \Delta_a$ and $\Delta_{hm} \Leftrightarrow \Delta_b$. Then $\Delta_{hl}=\Delta_a +\Delta_b > 0$. \\

$\bf Case~ 6$: $\Delta_{ml} \Leftrightarrow \Delta_a$ and $\Delta_{hl} \Leftrightarrow \Delta_b$. This case is excluded since it would imply $\Delta_{hm}=\Delta_b-\Delta_a < 0$, contradicting the definition given above. \\

Among these, $\bf Cases~3~ and~ 6$ are excluded based on the non-negativity constraints and the relation $\Delta_{hm} + \Delta_{ml} = \Delta_{hl}$. We now proceed to analyze the remaining four viable cases individually, as follows. \\

$\bf Case~1$ \\

In this case, let $\Delta_{hl} = \Delta_a$ and $\Delta_{hm} = \Delta_b$. Then,
\begin{eqnarray}
\Delta_{ml} = \Delta_a - \Delta_b = 2.4358 \times 10^{-3} {\rm eV}^2. 
\end{eqnarray} 
This implies that $\Delta_{hl} \sim \Delta_{ml} \gg \Delta_{hm}$, or equivalently, $m_h^2 \sim m_m^2 \gg m_l^2$. \\

Since $m^2_h -m^2_l= \Delta_a \approx m^2_m -m^2_l$, assuming $m^2_l$ is negligible,  
it follows that $\Delta_a$ must be very close to both $m^2_h$ and $m^2_m$.
Therefore, treating $\Delta_a$ as the midpoint bewteen $m_h^2$ and $m_m^2$, i.e., $(m^2_h +m^2_m) / 2  \approx  \Delta_a$, 
is a reasonable approximation.
Under this assumption, the predicted neutrino masses should be close to their actual values. \\

By combining  $(m^2_h +m^2_m) / 2 \approx \Delta_a$ with $(m^2_h -m^2_m) =\Delta_b$, we obtain the following:
\begin{eqnarray}
m_h &=& 5.04688 \times 10^{-2}~{\rm eV},~~~~~~m_m=4.97283 \times 10^{-2}~{\rm eV},~~~~~~ \nonumber \\ 
m_l &=& 6.09098 \times 10^{-3}~{\rm eV}, ~~~~~~g = 1.01489,\nonumber \\
g' &=& 8.16425, ~~~~~~~~~~~~~~~~~~~~~~g~ \cdot~ g'=8.28583. 
\end{eqnarray} 
The ratios $g$ and $g'$ align well with the constraints $m_h \sim m_m \gg m_l$. \\

$\bf Case~2$ \\

In this case, let $\Delta_{hl}=\Delta_a$ and $\Delta_{ml}=\Delta_b$. Then 
\begin{eqnarray}
\Delta_{hm}=\Delta_a -\Delta_b=2.4358 \times 10^{-3} {\rm  eV}^2. 
\end{eqnarray} 
This indicates that $\Delta_{hl} \sim \Delta_{hm} \gg \Delta_{ml}$, or equivalently, $m^2_h \gg m^2_m \sim m^2_l$. \\

Following a similar approach as in $\bf Case~ 1$, let $\Delta_b$ be the midpoint between $m^2_m$ and $m^2_l$, i.e,  $(m^2_m +m^2_l) / 2  \approx  \Delta_b$.
Combining this with $(m^2_m -m^2_l) =\Delta_b$, we obtain :
\begin{eqnarray}
m_h &=& 5.04688 \times 10^{-2}~{\rm eV},~~~~~~m_m=1.05499 \times 10^{-2}~{\rm eV},~~~~~~ \nonumber \\ 
m_l &=& 6.09098 \times 10^{-3}~{\rm eV},~~~~~~g = 4.78383,\nonumber \\
g' &=&1.73205, ~~~~~~~~~~~~~~~~~~~~~~g~ \cdot~ g'=8.28583. 
\end{eqnarray} 
The ratios $g$ and $g'$ do not align well with the expected mass hierarchy $m_h \gg m_m \sim m_l$. 
In particular, they are significantly smaller than those observed in the other three fermion types, 
and the value $g' = 1.73205$ suggests that $m_m$ is not particularly close to $m_l$. \\

$\bf Case~4$ \\

In this case, let $\Delta_{hm}=\Delta_a$ and $\Delta_{ml}=\Delta_b$. Then 
\begin{eqnarray}
\Delta_{hl}=\Delta_a +\Delta_b=2.5842 \times 10^{-3} {\rm eV}^2. 
\end{eqnarray}
That indicates $\Delta_{hl} \sim \Delta_{hm} \gg \Delta_{ml}$, or equivalently, $m^2_h \gg m^2_m \sim m^2_l$, similar to $\bf Case~ 2$. \\

Following the same approach as in $\bf Case~ 2$, let $\Delta_b$ be the midpoint between $m^2_m$ and $m^2_l$, i.e,  $(m^2_m +m^2_l) / 2  \approx  \Delta_b$.
Combining this with $(m^2_m -m^2_l) =\Delta_b$, we obtain :
\begin{eqnarray}
m_h &=& 5.11968 \times 10^{-2}~{\rm eV},~~~~~~m_m=1.05499 \times 10^{-2}~{\rm eV},~~~~~~ \nonumber \\ 
m_l &=& 6.09098 \times 10^{-3}~{\rm eV}, ~~~~~~g = 4.85301,\nonumber \\
g' &=& 1.73205, ~~~~~~~~~~~~~~~~~~~~~~g~ \cdot~ g'=8.405. 
\end{eqnarray} 
The masses of the lighter two neutrinos are identical to those in $\bf Case~ 2$; 
however, the heavier mass is slightly larger, with $m_h = 5.11968 \times 10^{-2}$ eV. \\

$\bf Case~5$ \\

In this case, let $\Delta_{ml}=\Delta_a$ and $\Delta_{hm}=\Delta_b$. Then 
\begin{eqnarray}
\Delta_{hl}=\Delta_a +\Delta_b=2.5842 \times 10^{-3} {\rm  eV}^2. 
\end{eqnarray}
That indicates $\Delta_{ml} \sim \Delta_{hl} \gg \Delta_{hm}$, or $m^2_h \sim m^2_m \gg m^2_l$, similar to $\bf Case~1$. \\

Following the same considerations as in $\bf Case~1$, $\Delta_a$ is treated as the midpoint between $m^2_h$ and $m^2_m$, i.e,  $(m^2_h +m^2_m) / 2 \approx \Delta_a$.
Combining this with $(m^2_h -m^2_m) =\Delta_b$, we obtain :
\begin{eqnarray}
m_h &=& 5.04688 \times 10^{-2}~{\rm eV},~~~~~~m_m=4.97283 \times 10^{-2}~{\rm eV},~~~~~~ \nonumber \\ 
m_l &=& 6.09098 i \times 10^{-3}~{\rm eV}, ~~~~~~ g = 1.01489,  \nonumber \\
g' &=& 8.16425 i, ~~~~~~~~~~~~~~~~~~~~~~g~ \cdot~ g' =8.28583 i. 
\end{eqnarray} 
The results obtained here are similar to those in $\bf Case~1$, except that $m_l$ is imaginary, which is unphysical. 
This indicates that the assumption $\frac{m_h^2 + m_m^2}{2} \approx \Delta_a$ must be rejected for this case. 
However, it lies close to the boundary of the physically allowed region,
 as will be confirmed through an alternative analytical approach in Subsection III-C and visualized in Fig. 9. \\

As a summary, the predicted value of $m_l \approx 6.09098 \times 10^{-3}$ eV remains consistent in three of the four cases. 
This value is notably different from the previous prediction of $m_1 =8.61 \times 10^{-3}$ eV, which corresponds the square root of $\Delta_b$ \cite{Esteban2020}. 
Alongside the earlier prediction of  $m_3=5.01 \times 10^{-2}$ eV, we now also predict various values for $m_h$ and the intermediate $m_m$. \\

There are  primarily  two groups of predictions for $m_m$: one suggests $m_m \approx 4.97283 \times 10^{-2}$ eV, assumed to be closer to $m_h$;
 while the other proposes $m_m \approx 1.05499 \times 10^{-2}$ eV, assumed to be closer to $m_l$.
In either case, the neutrino mass ratios are significantly smaller compared to those of the other three fermion types. 
The details are summerized in Table IV. \\

In $\bf Case~1~and ~5$, where $g=1.01489$, this ratio suggests that $m_h \approx m_m$,
 and the value $\Delta_a=2.51 \times 10^{-3}$ eV$^2$ may represent a combination of $\Delta_{hl}$ and $\Delta_{ml}$.
 If this interpretation is correct, future experiments with improved precision could potentially resolve the difference between these two MSDs. 
The predicted values presented here may serve as a useful reference for guiding the design of such experiments. \\

In contrast, for the other group with both $g' \approx 1.73205$, 
the difference between $m_m$ and $m_l$ is substantial, making them easier to distinguish compared to the previous group.
 However, such a large deviation has not been observed in current experiments, suggesting that these predictions may not be viable. \\

\begin{table}[tbp]
\centering
\begin{tabular}{l|llllll}
${\bf Case}$ &~~~ $m_h~{\bf (eV)}$ & ~~~$m_m~{\bf (eV)}$ & ~~~$m_l~{\bf (eV)}$ & ~~~$g$ & ~~~$g'$ & ~$\bf Physical$ \\ \hline
{\tt $ \begin{array}{ccccccc} {\bf 1} \\  {\bf 5}  \\  {\bf 2}  \\ {\bf 4}\end{array} $} &
{\tt $ \begin{array}{ccccccc} 5.05 \times 10^{-2} \\ 5.05  \times 10^{-2} \\ 5.05 \times 10^{-2} \\ 5.12  \times 10^{-2} \end{array} $} &
{\tt $ \begin{array}{ccccccc} 4.97 \times 10^{-2} \\ 4.97  \times 10^{-2} \\ 1.05 \times 10^{-2} \\ 1.05  \times 10^{-2} \end{array} $} &
$ \begin{array}{ccccccc} 6.09 \times 10^{-3} \\ 6.09i \times 10^{-3} \\ 6.09 \times 10^{-3} \\ 6.09 \times 10^{-3} \end{array}$ &
$ \begin{array}{ccccccc} 1.01 \\ 1.01 \\ 4.78 \\ 4.85 \end{array} $ &
$ \begin{array}{ccccccc} 8.16 \\ 8.16i \\ 1.73 \\ 1.73 \end{array} $ &
$ \begin{array}{ccccccc} {\rm Yes} \\ {\rm No} \\ {\rm Possibly} \\ {\rm Possibly} \end{array} $ 
\\ \hline
\end{tabular}

\caption{\label{tab:i} 
Case 5 yields an imaginary value for $m_l$, making it unphysical and excluding it from further consideration. The remaining three cases are noteworthy for future investigations.}
\end{table}

\subsection{The Mass Hierarchies and Leptogenesis }\label{sect3B}

Following the discussions in the previous subsection, the author extends these definitions to all four fermion types. 
The following mass ratios are obtained: \\

$\bf 1. ~For ~up-type ~quarks$
\begin{eqnarray}
g_{(u)} &\equiv &{m_t \over m_c} ~>~ {(172.0-0.9-1.3) \over (1.27+0.07)} \sim 126.7, ~~~~ \nonumber \\ 
g'_{(u)}  &\equiv & {m_c \over m_u} ~>~ {(1.27-0.09) \over 0.0033} \sim 357.6 .
\end{eqnarray} 

$\bf 2. ~For ~down-type ~quarks$
\begin{eqnarray}
g_{(d)}   &\equiv & {m_b \over m_s} ~>~ {(4.19-0.06) \over (0.101+0.029)} \sim 31.77, ~~~~ \nonumber \\ 
g'_{(d)}   &\equiv &{m_s \over m_d} ~>~ {(0.101-0.021) \over 0.0058} \sim 13.79.
\end{eqnarray} 

$\bf 3. ~For ~charged ~leptons$
\begin{eqnarray}
g_{(\ell)}  &\equiv & {m_{\tau} \over m_{\mu}} ~>~ {1777 \over 105.7} \sim 16.81, ~~~~ \nonumber \\ 
g'_{(\ell)} &\equiv & {m_{\mu} \over m_e} ~>~ {105.7 \over 0.511} \sim 206.8.
\end{eqnarray} 
Note: In the equations above, the maximum of the masses in the denominator and the minimum in the numerator are chosen to ensure the "$>$" signs always hold true.
Among these six ratios, the smallest one is $g'_{(d)} \approx 13.79$. This value is much larger than the comparable ratios obtained for neutrinos, as shown below.\\ 

$\bf 4. ~For ~neutrinos$
The candidate ratio sets are: \\
 
$\bf Case ~1$, $g_{(\nu)} = 1.01489$, $g'_{(\nu)}=8.16425$, and $g_{(\nu)} \cdot g'_{(\nu)} =8.28583$. \\ 

$\bf Case ~2$, $g_{(\nu)} = 4.78383$, $g'_{(\nu)}=1.73205$, and $g_{(\nu)} \cdot g'_{(\nu)}=8.28583$. \\ 

$\bf Case ~4$, $g_{(\nu)} = 4.85301$, $g'_{(\nu)}=1.73205$, and  $g_{(\nu)} \cdot g'_{(\nu)} =8.40500$. \\ 

Considering the mass ratios in the quark sector and in charged leptons, $g'_{(d)} \equiv {m_s \over m_d} \approx 13.79 $ is the smallest  among these three fermion types. 
 The difference between $(m^2_s-m^2_d)$ and $m^2_s$ is only about ${1 \over g'^2_{(d)}} \approx {1 \over 190.2}$ of $m^2_s$. 
 It is therefore reasonable to ignore the mass of the lighter fermion in such a MSD.
However, the ratios $g_{(\nu)}$ and $g'_{(\nu)}$ obtained in Subsection III-A do not justify such approximations in any of the neutrino cases. \\

In each of the remaining three viable cases, the product $g \cdot g' \equiv {m_h \over m_l}$ range between 8.28583 and 8.40500, which are significantly smaller than the corresponding ratios in the other three fermion types.
These values are clearly too small to disregard any $m_l$ in the subsequent derivations for neutrinos. \\ 

In the quark sector, Jarlskog suggested a measure for the strength of CP violation \cite{Jarlskog1985}:
\begin{eqnarray}
\Delta_{CP} &=& {\rm Im~ Det} [m_u m_u^{\dagger} , m_d m_d^{\dagger} ] ~T^{-12} \nonumber \\
&=& {\it J}~ \prod_{\scriptstyle i < j} (m_{u,i}^2 - m_{u,j}^2 ) \prod_{\scriptstyle i < j}(m_{d,i}^2 - m_{d,j}^2 )~ T^{-12} \nonumber \\
&=&~ {\it J}~ \Delta m^2_{(u)} ~\Delta m^2_{(d)} ~T^{-12}~,
\end{eqnarray}
where $J$ is the Jarlskog invariant, $T\approx$ 100 GeV is the temperature of the electroweak phase transition, and $m^2$ represents squares of quark masses. \\

In the last line of Eq.(63), $ \Delta m^2_{(u)}$ and $ \Delta m^2_{(d)}$ are the products of three MSDs in the up- and down-type quarks, defined as:
\begin{eqnarray}
\Delta m_{(u)}^2 &=& (m_t^2 - m_c^2 ) (m_c^2 - m_u^2 ) (m_u^2 - m_t^2 ) \nonumber \\
&=& -(m_t^2 - m_c^2 ) (m_c^2 - m_u^2 ) (m_t^2 - m_u^2 ) ~<~0, \\
\Delta m_{(d)}^2 &=& (m_b^2 - m_s^2 ) (m_s^2 - m_d^2 ) (m_d^2 - m_b^2 ) \nonumber \\ 
&=& -(m_b^2 - m_s^2 ) (m_s^2 - m_d^2 ) (m_b^2 - m_d^2 )~<~0,
\end{eqnarray} 
respectively. \\

In the lepton sector, the maximally allowed CP-violating Jarlskog invariant was estimated to be \cite{Gonzalez2014}:
\begin{eqnarray}
J^{max}_{(l)}~=~0.033 \pm 0.010~\pm (0.027).
\end{eqnarray} \\

In the expression for CP violation in Eq.(63), six MSDs appear in the quark sector: three from up-type quarks and three from down-type quarks. Similarly, there should be three MSDs from charged leptons and three from neutrinos in the lepton sector.
From recent global analyses of three-flavor neutrino oscillations, the neutrino MSDs are given by:
\begin{eqnarray}
\Delta m^2_{31} &=& ~~~2.517^{+0.026}_{-0.028} \cdot ~10^{-3} ~ {\rm eV}^2,~~( {\rm NO}) \\
\Delta m^2_{32} &=& -2.498^{+0.026}_{-0.028} \cdot ~10^{-3} ~ {\rm eV}^2,~~( {\rm IO}) \\
\Delta m^2_{21} &=& ~~~7.42^{+0.21}_{-0.20} ~~~~\cdot ~10^{-5} ~ {\rm eV}^2, 
\end{eqnarray}
where $\Delta m^2_{ij} = m^2_i -m^2_j$ denotes the MSD of two neutrinos, 
and NO (IO) is the abbreviation for Normal ordering (Inverted ordering), defined by $m_1 < m_2 < m_3$ ($m_3 < m_1 < m_2$). 
However, only two of the MSDs are experimentaly obtained. \\

In general, two given values are insufficient to analytically determine three unknowns. 
However, in the case of mass-squared differences (MSDs), the third MSD can be determined unambiguously due to the constraint:
\begin{eqnarray}
\Delta_{hm}+ \Delta_{ml} \equiv (m^2_h -\cancel{m^2_m})+(\cancel{m^2_m}-m^2_l)=(m^2_h -m^2_l) \equiv \Delta_{hl}.
\end{eqnarray}
This identity ensures that any two of the MSDs uniquely determine the third. 
The remaining question is how the two experimentally measured quantities, $\Delta_a$ and $\Delta_b$, correspond to the theoretical MSDs: $\Delta_{hl}$, $\Delta_{hm}$, and $\Delta_{ml}$. \\

As discussed in Subsection III-A, there are six possible correspondences between $\Delta_a$ and $ \Delta_b$, and three $\Delta_{ij}$.
Among these, only four candidates are logically self-consistent: \\

$\bf Case~ 1:$ Let $\Delta_a~=~\Delta_{hl}~=~(m^2_h -m^2_l)$ and $\Delta_b~=~\Delta_{hm}~=$ $(m^2_h -m^2_m)$, then $\Delta_{ml}~=~(m^2_m -m^2_l)~=~\Delta_a-\Delta_b=2.4358~\cdot ~10^{-3}$ eV$^2$. \\

$\bf Case~ 2:$  Let $\Delta_a~=~\Delta_{hl}~=~(m^2_h -m^2_l)$ and $\Delta_b~=~\Delta_{ml}~=$ $(m^2_m -m^2_l)$, then $\Delta_{hm}~=~(m^2_h-m^2_m)=\Delta_a -\Delta_b =2.4358~\cdot ~10^{-3}$ eV$^2$. \\

$\bf Case~ 4:$ Let $\Delta_a~=~\Delta_{hm}~=~(m^2_h -m^2_m)$ and $\Delta_b~=~\Delta_{ml}~=$ $(m^2_m -m^2_l)$, then $\Delta_{hl}~=~(m^2_h-m^2_l)=\Delta_a +\Delta_b=2.5842~\cdot ~10^{-3}$ eV$^2$. \\

$\bf Case~ 5:$  Let $\Delta_a~=~\Delta_{ml}~=~(m^2_m -m^2_l)$ and $\Delta_b~=~\Delta_{hm}~=$ $(m^2_h -m^2_m)$, then $\Delta_{hl}~=~(m^2_h -m^2_l)=\Delta_a +\Delta_b=2.5842~\cdot ~10^{-3}$ eV$^2$. \\

There is a particularly interesting quantity, $\Delta m^2_{(\nu)}$, the product of three MSDs for neutrinos, defined by:
\begin{eqnarray}
\Delta m^2_{(\nu)} &\equiv & (m_h^2 - m_l^2 )_{(\nu)}~ (m_h^2 - m_m^2 )_{(\nu)}~ (m_m^2 - m_l^2 )_{(\nu)} \nonumber \\
&=& (\Delta_{hl} \cdot \Delta_{hm} \cdot \Delta_{ml})_{(\nu)} = 2 \gamma_{(\nu)}~ (\beta^2_{(\nu)} -\gamma^2_{(\nu)}),
\end{eqnarray} 
which is almost the same in all cases.
Besides, it indicates that $\Delta m^2_{(\nu)}$ is independent of the parameter $\alpha_{(\nu)}$. \\

Upon substituting the results obtained in the neutrino sector into Eq.(71), the following outcomes are derived: \\

In $\bf Cases~ 1 ~and ~2:$  
\begin{eqnarray}
 \vert \Delta m_{(\nu)}^2  \vert = \Delta_a ~\Delta_b ~(\Delta_a -\Delta_b)~=~4.8129 \times 10^{-64}~{\rm GeV}^6 .
\end{eqnarray} 

In $\bf Cases~ 4 ~and ~5:$
\begin{eqnarray}
 \vert \Delta m_{(\nu)}^2 \vert = \Delta_a ~\Delta_b ~(\Delta_a +\Delta_b)~=~4.5365 \times 10^{-64}~{\rm GeV}^6 . 
\end{eqnarray} 
These products of neutrino MSDs are remarkably similar regardless of how $\Delta_a$ and $\Delta_b$ correspond to the three $\Delta_{ij}$. 
However, the $\Delta m_{(\nu)}^2$ values are dramatically smaller than the similar quantities in the other three fermion types: \\

$\bf For~ up-type~ quarks:$
\begin{eqnarray}
\vert \Delta m^2_{(u)} \vert \approx 1.463~\times 10^9 ~{\rm GeV}^6, 
\end{eqnarray} 
(Using $m_t$= 173.21 GeV, $m_c$= 1.275 GeV, and $m_u$= 0.0023 GeV). \\

$\bf For~ down-type~ quarks:$
\begin{eqnarray}
\vert \Delta m_{(d)}^2 \vert \approx 2.747 ~{\rm GeV}^6,
\end{eqnarray} 
(Using $m_b$= 4.180 GeV, $m_s$= 0.095 GeV, and $m_d$= 0.0048 GeV). \\

$\bf For~charge~ leptons:$
\begin{eqnarray}
\vert \Delta m^2_{(\ell)} \vert \approx 0.1107 ~{\rm GeV}^6,
\end{eqnarray} 
(Using $m_{\tau}$= 1.7768 GeV, $m_{\mu}$= 0.1056 GeV, and $m_e$= 0.000511 GeV). \\

The MSD products for the other three fermion types are at least 62 orders of magnitude larger than that of neutrinos. 
This vast hierarchy remains an unexplained mystery in physics. \\

By substituting the MSD products for the quark sector into the CPV measure Eq.(63), we obtain:
\begin{eqnarray}
\vert \Delta_{CP(q)} \vert \approx J_{(q)} \cdot 4.019 \times 10^9 ~{\rm GeV}^6,
\end{eqnarray} 
while for the lepton sector:
\begin{eqnarray}
\vert \Delta_{CP(l)} \vert \approx J_{(l)} \cdot  (^{5.3279}_{5.0219}) \times 10^{-65} ~{\rm GeV}^6.
\end{eqnarray} 

Taking the Jarlskog invariant in the quark sector as $J_{(q)} = 3.0 \times 10^{-5}$ \cite{Zyla2020}  and the maximally allowed CP-violating Jarlskog invariant in the lepton sector, $J_{(l)} \approx 0.033$ \cite{Gonzalez2014},
 the CPV measure in the quark sector is still at least 71 orders of magnitude greater than that in the lepton sector. 
This stark difference suggests that leptogenesis in the electroweak standard model is negligible in comparison to baryogenesis in our current universe. \\

\subsection{An Alternative Approach to Studying Neutrino Masses}\label{sect III-C}

This subsection presents an alternative analysis of neutrino masses, 
investigating their dependence on the mass ratio $g\equiv m_h / m_m$ and the corresponding constraints imposed by the model. \\

In $\bf Case~ 1$ of Subsection III-A, the following relationships are observed:
\begin{eqnarray}
{\Delta_a \over \Delta_b} &=& 33. 8275={ {(g^2 g'^2-1) \cancel{m^2_l}} \over {{(g^2 g'^2-g'^2)} \cancel{m^2_l} }}, \\
g' &= & \sqrt{1 \over {33.8275-32.8275 g^2}}, \\ 
m_l &=& \sqrt{ \Delta_b \over {g'^2 (g^2-1)}}. 
\end{eqnarray} 
Fig. 2 illustrates the variation of $g'$ with respect to $g$, showing that $g'$ increases sharply toward infinity as $g$ approaches 1.01512. 
This divergence occurs as the denominator of Eq.(80) approaches zero. \\

Furthermore, Fig. 3 presents the variation of $m_h$, $m_m$, and $m_l$ with respect to $g$. 
In this figure, $m_h \approx m_m \approx m_l$ when $g$ approaches 1, 
but $m_l$ diverges from the other two as $g$ increases and decreases rapidly to zero as $g$ approaches 1.01512. 
Beyond that point, $m_h$ and $m_m$ become negative and $m_l$ becomes imaginary, which are obviously unphysical. \\

Consequently, physically meaningful neutrino masses that satisfy $m_h > m_m > m_l > 0$ are only allowed within a very narrow range $1 < g < 1.01512$. 
In Fig. 2, two reference points are plotted: a blue point at $(g, g') = (1.01489, 8.16425)$, where $g \cdot g' \approx 8.28583$,
corresponds to the results obtained in Eq.(53); 
and a green point at $(g, g') = (1.01467, 5.79514)$ is obtained by substituting the value $m_1 = 8.61 \times 10^{-3}$ eV predicted in \cite{Esteban2020} into Eqs. (81). 
With these reference points established, the range between these two points will be a primary focus of our attention in the future.\\

\begin{figure}
\includegraphics{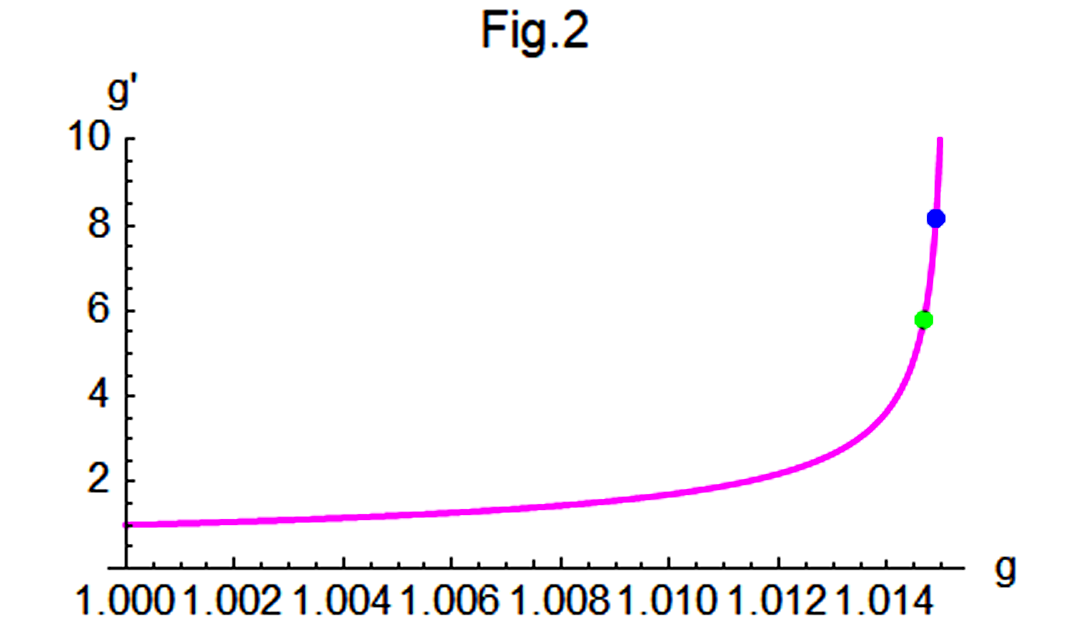}
\label{fig:petit3leaks3D}
\caption{ The variation of $g'$ with $g$ reveals that $g'$ increases sharply toward infinity as $g$ approaches 1.0151. 
Consequently, the self-consistent range in this case is restricted to a very narrow interval, $1 < g < 1.01512$. 
For reference, two points are marked in the figure: the blue point at ($g,~g'$) = (1.01489, 8.16425) represents the result obtained in Eq.(53) of Subsection III-A, 
while the green point at ($g,~g'$) = (1.01467, 5.79514) is obtained by substituting the predicted value $m_1 = 8.61 \times 10^{-3}$ eV from \cite{Esteban2020} into Eq.(81). 
The region between these two points is an area that future experimental designs should pay closer attention to. }
\label{figure2}
\end{figure}

\begin{figure}
\includegraphics{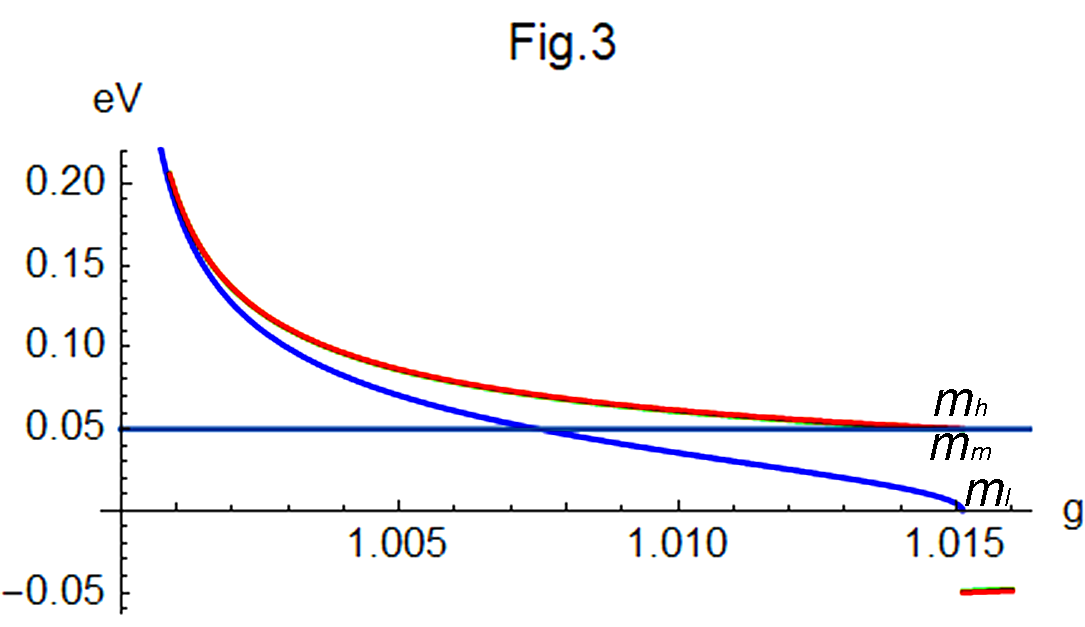}
\label{fig:petit3leaks3D}
\caption{ The variations of $m_h$, $m_m$, and $m_l$ with $g$ reveal that the masses of the three neutrinos are nearly identical when $g$ is close to 1. 
As $g$ increases, $m_l$ gradually deviates from the other two and approaches zero as $g$ approaches 1.01512, while $m_m$ remains very close to $m_h$.
Beyond this point, the masses become unphysical. }
\label{figure3}
\end{figure}

In $\bf Case~ 2$ of Subsection III-A, the following relationships are observed:
\begin{eqnarray}
{\Delta_a \over \Delta_b} &=& 33. 8275={ {(g^2 g'^2-1) \cancel{m^2_l}} \over {{(g'^2-1)} \cancel{m^2_l} }}, \\
g' &= & \sqrt{32.8275 \over {33.8275-g^2}}. \\
m_l &= & \sqrt{\Delta_b \over {g'^2-1}}. 
\end{eqnarray} 
Fig. 4 illustrates the variation of $g'$ with respect to $g$, showing that $g'$ increases sharply toward infinity as $g$ approaches 5.81614.
Such a divergence occurs as the denominator of Eq.(83) approaches zero. \\

Furthermore, Fig. 5 presents the variation of $m_h$, $m_m$, and $m_l$ with respect to $g$. 
In this figure, $m_h \approx m_m \approx m_l$ as $g$ approaches 1;
however, $m_h$ diverges from the other two as $g$ increases. 
As $g$ further increases, $m_h$ approaches a constant value of approximately $5.01 \times 10^{-2}$ eV, while $m_m$ and $m_l$ remain very close to each other, decreasing gradually until $g$ approaches 5.81614.
At this point, where $g^2 = {\Delta_a \over \Delta_b}$, $m_l$ drops to zero. 
Beyond that point, unphysical negative and imaginary neutrino masses emerge. \\

Consequently, physically meaningful neutrino masses that satisfy $m_h > m_m > m_l > 0$ occur only within the range $1 < g < 5.81614$.
In Fig. 4, two reference points are plotted: a blue point at $(g, g') = (4.78383, 1.73205)$, 
which corresponds to the results from Eq.(55) and aligns well with the curve; 
and a green point at $(g, g') = (4.17388, 1.41454)$, obtained by substituting the predicted value $m_1 = 8.61 \times 10^{-3}$ eV into Eq.(84). 
The range between these two points will be a primary focus of our attention in the future. \\

\begin{figure}
\includegraphics{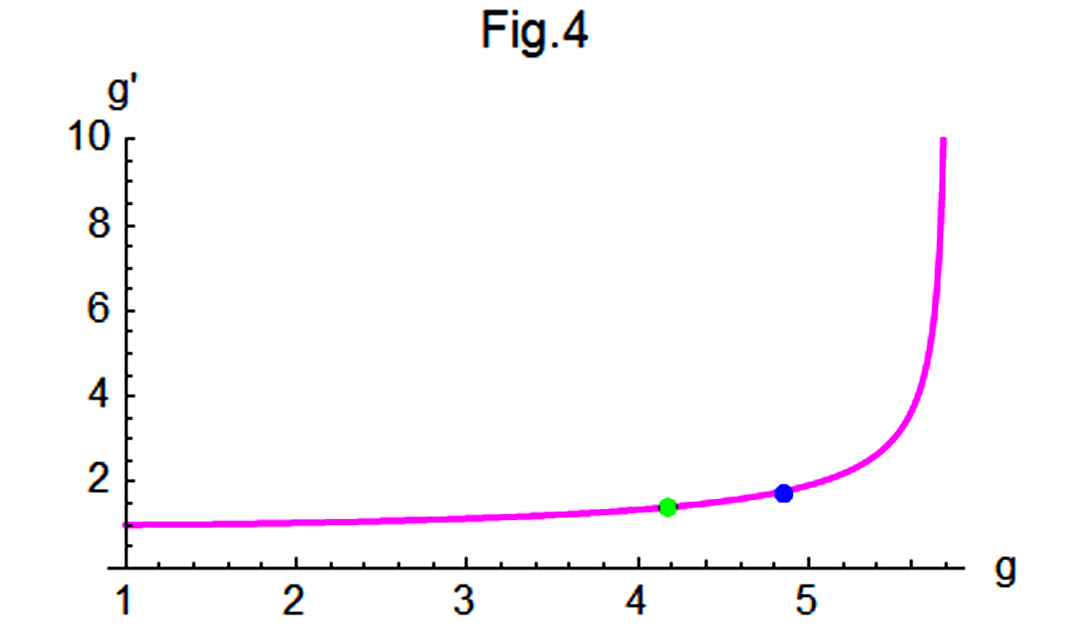}
\label{fig:petit3leaks3D}
\caption{
The variation of $g'$ with $g$ reveals that $g'$ increases sharply toward infinity as $g$ approaches 5.81614. 
Consequently, the self-consistent range for this case lies within $1 < g < 5.81614$.
For reference, two points are marked in the figure: the blue point at ($g,~g'$) = (4.78383, 1.73205) represents the result obtained in Eq.(55) of Subsection III-A, 
while the green point at ($g,~g'$) = (4.17388, 1.41454) is obtained by substituting the predicted value $m_1 = 8.61 \times 10^{-3}$ eV from \cite{Esteban2020} into Eq.(84). 
The region between these two points is an area that future experimental designs should pay closer attention to.
}
\label{figure4}
\end{figure}

\begin{figure}
\includegraphics{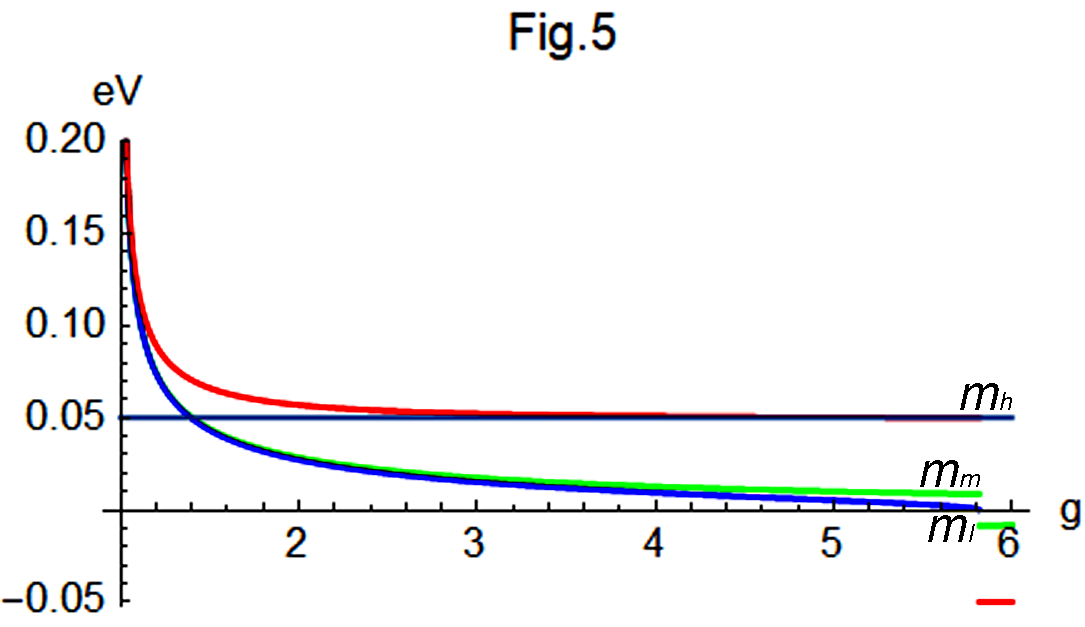}
\label{fig:petit3leaks3D}
\caption{ The variations of $m_h$, $m_m$, and $m_l$ with $g$ show that the masses of the three neutrinos are nearly identical when $g$ is close to 1. 
As $g$ increases, $m_h$ begins to deviate from the other two, while $m_l$ drops to zero as $g$ approaches 5.81614. 
Beyond this point, the masses become unphysical. }
\label{figure5}
\end{figure}

In $\bf Case~ 4$ of Subsection III-A, the following relationships are observed:
\begin{eqnarray}
{\Delta_a \over \Delta_b} &=& 33. 8275={ {g'^2 (g^2 -1) \cancel{m^2_l}} \over {{(g'^2-1)} \cancel{m^2_l} }}, \\
g' &= & \sqrt{33.8275 \over {34.8275-g^2} }, \\
m_l &= & \sqrt{\Delta_b \over {g'^2-1} }. 
\end{eqnarray} 
Fig. 6 illustrates the variation of $g'$ with respect to $g$, showing that $g'$ increases sharply to infinity as $g$ approaches 1.01467, 
where the denominator of Eq.(86) approaches zero. 
This result is very similar to that obtained in $\bf Case~2 $, but with slight differences. \\

Furthermore, Fig.7 presents the variation of $m_h$, $m_m$, and $m_l$ with respect to $g$.
In this figure, the three masses $m_h \approx m_m \approx m_l$ converge as $g$ approaches 1, 
but $m_h$ diverges from the other two as $g$ increases. 
As $g$ increases further, $m_h$ approaches a constant value of approximately $5.01 \times 10^{-2}$ eV, 
while $m_m$ and $m_l$ remain very close to each other and decrease slowly until $g$ approaches 5.90148,
at which point $g^2 = {\Delta_a \over \Delta_b}$ and $m_l$ drops to zero sharply. 
Beyond that point, unphysical negative and imaginary neutrino masses emerge. \\

Consequently, physically meaningful neutrino masses satisfying $m_h > m_m > m_l > 0$ occur only within the range $1 < g < 5.90148$.
In Fig. 6, two reference points are plotted: a blue point at $(g, g') = (4.85301, 1.73205)$, 
which corresponds to the results from Eq.(57) and aligns well with the curve; 
and a green point at $(g, g') = (4.23338, 1.41454)$, obtained by substituting the predicted value $m_1 = 8.61 \times 10^{-3}$ eV \cite{Esteban2020} into Eqs. (87).\\

\begin{figure}
\includegraphics{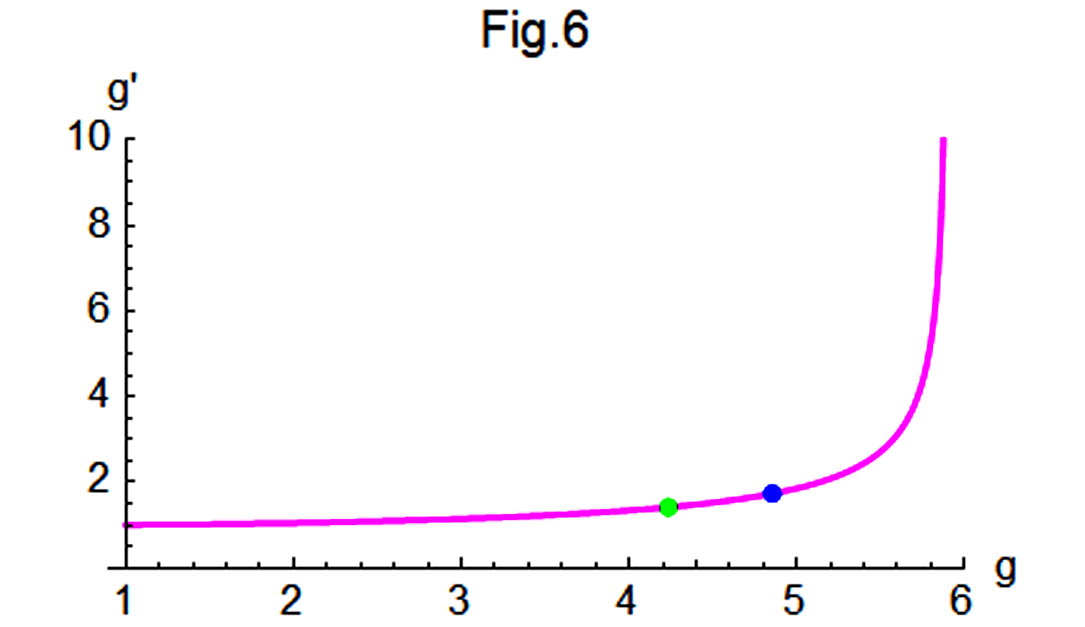}
\label{fig:petit3leaks3D}
\caption{ The variation of $g'$ with $g$ reveals that $g'$ increases sharpdly towards infinity as $g$ approaches 5.90148. 
Consequently, the self-consistent range for this case lies within $1 < g < 5.90814$.
For reference, two points are marked in the figure: the blue point at ($g,~g'$) = (4.85301, 1.73205) represents the result obtained in Eq.(57) of Subsection III-A, 
while the green point at ($g,~g'$) = (4.23338, 1.41454) is obtained by substituting the predicted value $m_1 = 8.61 \times 10^{-3}$ eV from \cite{Esteban2020} into Eq.(87).
The region between these two points is an area that future experimental designs should pay closer attention to. }
\label{figure6}
\end{figure}

\begin{figure}
\includegraphics{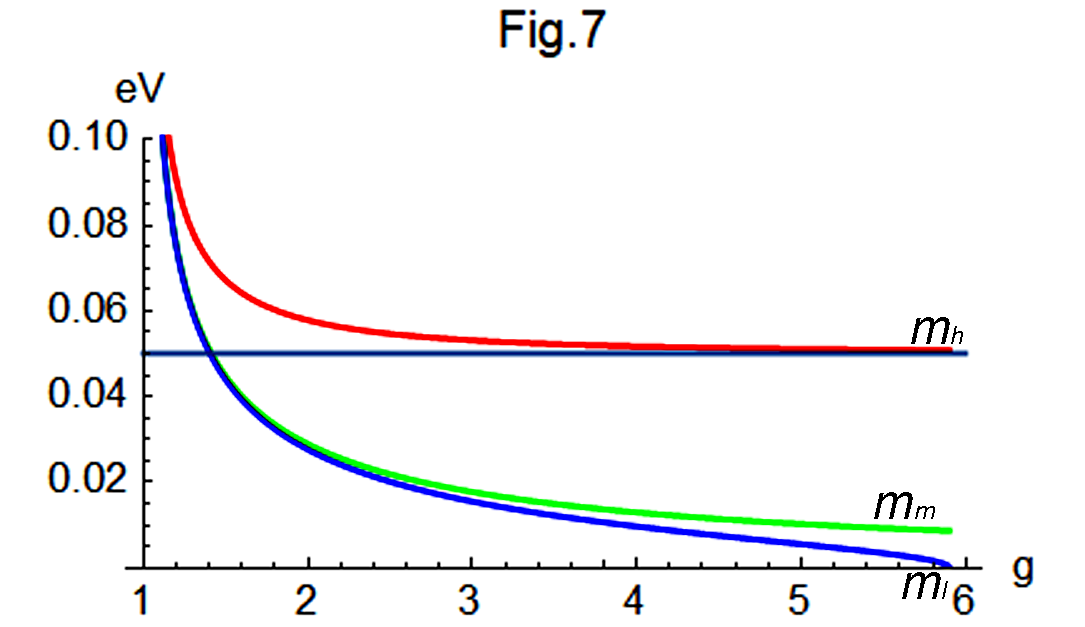}
\label{fig:petit3leaks3D}
\caption{ The variations of $m_h$, $m_m$, and $m_l$ with $g$ indicate that the masses of the three neutrinos are nearly identical when $g$ is close to 1. 
As $g$ increases, $m_h$ starts to deviate from the other two, and $m_l$ drops to zero as $g$ approaches 5.90148. 
Beyond this point, the masses become unphysical. }
\label{figure7}
\end{figure}

In $\bf Case~5 $ of Subsection III-A, the following relationships are observed:
\begin{eqnarray}
{\Delta_a \over \Delta_b} &=& 33. 8275={ { (g'^2 -1) \cancel{m^2_l}} \over {{g'^2 (g^2-1)} \cancel{m^2_l} }}, \\
g' &= & \sqrt{1 \over {34.8275-33.8275 g^2} }, \\
m_l &= & \sqrt{\Delta_a \over {g'^2-1} }. 
\end{eqnarray} 
Fig. 8 illustrates the variation of $g'$ with respect to $g$, showing that $g'$ increases sharply to infinity as $g$ approaches 1.01467, where the denominator of Eq.(89) approaches zero. 
This result is very similar to that obtained in $\bf Case~1 $ of Subsection III-A, but with slight differences. \\

Furthermore, Fig. 9 presents the variation of $m_h$, $m_m$, and $m_l$ with respect to $g$.
In this figure, $m_h \approx m_m \approx m_l$ when $g$ approaches 1, but $m_l$ begins to diverge from the other two as $g$ increases. 
While $m_h$ and $m_m$ remain very close to each other, both soon approach approximately $5.01 \times 10^{-2}$ eV as $g$ increases. 
In contrast, $m_l$ decreases rapidly to zero when $g$ approaches 1.01467, at which point $g^2 = {\Delta_a \over \Delta_b}$.
Beyond this point, unphysical negative and imaginary neutrino masses appear. \\

Consequently, physically reasonable neutrino masses satisfying $m_h > m_m > m_l > 0$ arise only within a very narrow range $1 < g < 1.01467$. 
Two reference points are plotted in Fig. 9: a blue point at $(g, g') = (1.01489, 8.16425)$, corresponding to the results obtained in Eq.(59); 
and a green point at $(g, g') = (1.01426, 5.93918)$, obtained by substituting the previously predicted value $m_1 = 8.61 \times 10^{-3}$ eV into Eq.(90). 
Unlike the previous cases, the blue point lies slightly to the right of the curve, and the resulting imaginary value of $m_l$ in Eq.(59) suggests that the assumption of ${{m^2_h +m^2_m} \over 2} \approx \Delta_a$ is not valid in this region.
However, this discrepancy does not rule out the scenario; rather, it implies that the theoretically allowed upper bound of $g$ is more tightly constrained. 
Therefore, the region of interest should be further restricted to the narrower interval  $1.01426  < g < 1.01467$. \\

\begin{figure}
\includegraphics{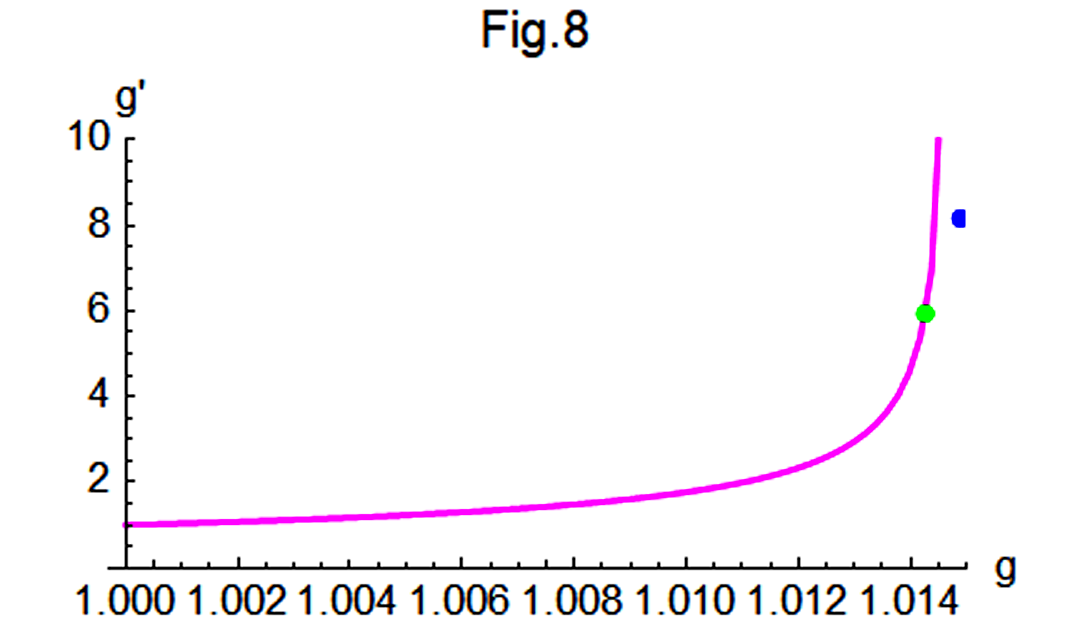}
\label{fig:petit3leaks3D}
\caption{ The variation of $g'$ with $g$ reveals that $g'$ increases sharply toward infinity as $g$ approaches 1.01467. 
Consequently, the self-consistent range in this case is restricted to a very narrow interval, $1 < g < 1.01467$. 
For reference, two points are marked in the figure: the blue point at ($g,~g'$) = (1.01489, 8.16425) represents the result obtained in Eq.(59) of Subsection III-A, 
while the green point at ($g,~g'$) = (1.01426, 5.93918) is obtained by substituting the predicted value $m_1 = 8.61 \times 10^{-3}$ eV from \cite{Esteban2020} into Eq.(90).
The region between these two points is an area that future experimental designs should pay closer attention to. }
\label{figure8}
\end{figure}

\begin{figure}
\includegraphics{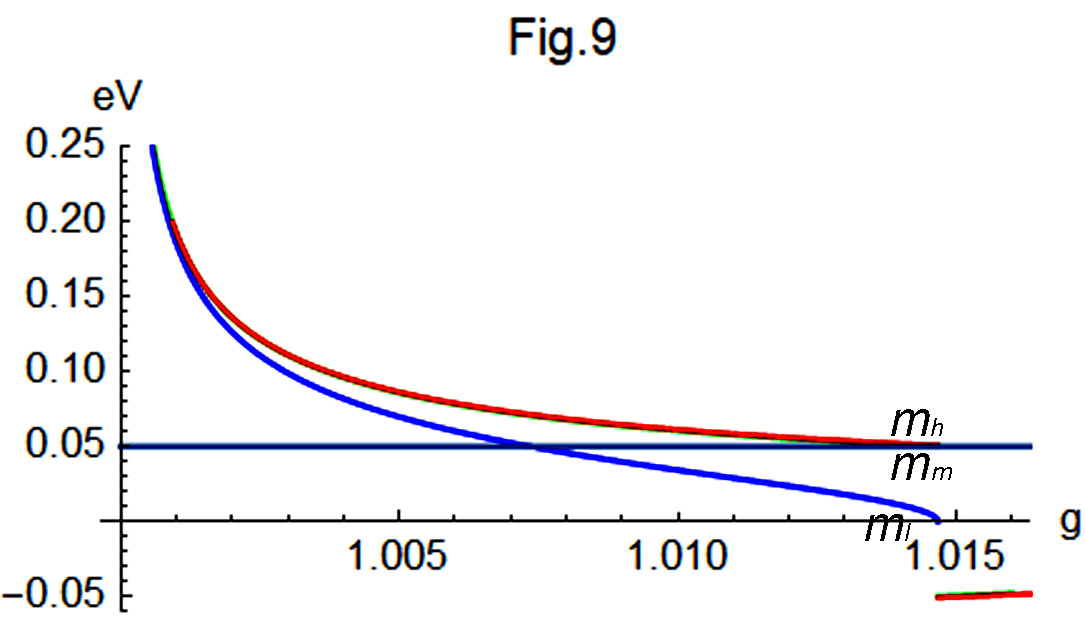}
\label{fig:petit3leaks3D}
\caption{ The variations of $m_h$, $m_m$, and $m_l$ with $g$ reveal that the masses of the three neutrinos are nearly identical when $g$ is close to 1. 
As $g$ increases, $m_h$ begins to deviate from the other two, while $m_l$ drops to zero as $g$ approaches 1.01467. 
Beyond this point, the masses become unphysical. }
\label{figure9}
\end{figure}

{\bf Section~Summary} \\

The findings from all three subsections can summarized as follows:
In this section, various approaches to investigate the masses of neutrinos are explored. 
In Subsection III-A, two of the six possible ways to match the two experimentally given values, $\Delta_a$ and $\Delta_b$, 
with the three theoretically defined MSDs $\Delta_{hm}, \Delta_{ml}$, and $\Delta_{hl}$ are  excluded due to inconsistencies. 
Among the remaining four viable cases, two exhibit $m_m \approx m_h$, while the other two exhibit $m_m \approx m_l$.
Accordingly, we tested the midpoint $\Delta_a \approx {{m^2_h +m^2_m}\over 2}$ for cases where $m_h \approx m_m$ and $\Delta_b \approx {{m^2_m +m^2_l}\over 2}$ for cases where $m_m \approx m_l$. \\

As a result, $m_l$ is consistently predicted to be 6.09098 $\times 10^{-3}$ eV in all cases, differing from previous analyses in \cite{Esteban2020}. 
The predictions for $m_h$ converge around $5.01 \times 10^{-2}$ eV in all cases. 
However, predictions for $m_m$ fall into two groups: \\

In $\bf Cases~ 1~ and~ 5$, $m_m$ = 4.97283 $\times 10^{-2}$ eV is closer to $m_h$. 

In $\bf Cases~ 2~ and~ 4$, $m_m$ = 1.05499 $\times 10^{-2}$ eV is closer to $m_l$. \\

In Subsection III-B, through analysis of the MSDs, all four viable cases predict an almost identical value for $\Delta m^2_{(\nu)} = ^{4.8129}_{4.5365} \times 10^{-64}$ GeV$^6$, 
which is approximately 62 orders of magnitude smaller than the smallest $\Delta m^2_{(\ell)} \approx 0.1107$ GeV$^6$ of the charged leptons. 
With all four MSD products determined, the Jarlskog measure of CPV is also calculated, 
revealing that leptogenesis in the Standard Electroweak Model is around 71 orders of magnitude smaller than baryogenesis in the current universe. 
This underscores the need for Beyond Standard Model (BSM) physics if we expect leptogenesis to play a significant role in resolving the Baryon Asymmetry of the Universe. \\

In Subsection III-C, a more  comprehensive analysis on the neutrino masses is provided.
The self-consistent ranges of $g$ and $g'$ for each case are studied, and the variations of $g'$, $m_h$, $m_m$, and $m_l$ with respect to $g$ are plotted.
The results can be summarized as follows: \\

1. Two cases (1 and 5) suggest that $m_h \sim m_m \approx 5.01 \times 10^{-2}$ eV and $m_l \approx 6.09098 \times 10^{-3}$ eV, with $g$ constrained to very narrow ranges: 
\begin{eqnarray}
1 &<& g < 1.01512~~~~~~~~~~{\rm in}~ {\bf Case~ 1}, \\
1 &<& g < 1.01467~~~~~~~~~~{\rm in}~ {\bf Case~ 5}. 
\end{eqnarray} 

2. The other two cases (2 and 4) indicate wider ranges: 
\begin{eqnarray}
1 &<& g < 5.81614~~~~~~~~~~{\rm in}~ {\bf Case~ 2}, \\
1 &<& g < 5.90148~~~~~~~~~~{\rm in}~ {\bf Case~ 4}. 
\end{eqnarray} 

In addition, two reference points are plotted in Figs. 2, 4, 6, and 8, respectively, to illustrate the results previously analyzed. 
The intervals between each pair of points highlight the significant ranges of the variable $g$ in the corresponding cases:
\begin{eqnarray}
1.01467 &<& g < 1.01489~~~~~~~~~~{\rm in}~ {\bf Case~ 1}, \\
4.17388 &<& g < 4.78383~~~~~~~~~~{\rm in}~ {\bf Case~ 2}, \\
4.23338 &<& g < 4.85301~~~~~~~~~~{\rm in}~ {\bf Case~ 4}, \\
1.01426 &<& g < 1.01467~~~~~~~~~~{\rm in}~ {\bf Case~ 5}. 
\end{eqnarray}
These ranges correspond to the following intervals of the lightest neutrino mass $m_l$: 
\begin{eqnarray}
8.64611 \times 10^{-3}{\rm eV} &>& m_l > 6.11487 \times 10^{-3}{\rm eV}, \\
8.60999 \times 10^{-3}{\rm eV} &>& m_l > 6.09096 \times 10^{-3}{\rm eV}, \\
8.60999 \times 10^{-3}{\rm eV} &>& m_l > 6.09096 \times 10^{-3}{\rm eV}, \\
8.55943 \times 10^{-3}{\rm eV} &>& m_l > 7.45159 \times 10^{-4}{\rm eV}, 
\end{eqnarray}
respectively.
These intervals represent parameter regions that warrant closer attention in future analyses. 
Notably, the lower bound 7.45159 $\times 10^{-4}$ eV in Eq.(102) (\textbf{Case 5}) extends to a value significantly lower than those in the other three cases.  \\

\section{Conclusions and Discussions}

In this article, the neutrino mass spectrum has been analyzed within an analytically solvable CP-Violating Standard Model (CPVSM). 
Using two experimentally measured mass squared differences (MSDs) along with the fundamental relationship among the three MSDs defined in Eq.(2) by $\Delta_{hm} + \Delta_{ml} \equiv \Delta_{hl}$,
 we have successfully determined the third MSD in various cases.
 This approach enables the calculation of the MSD product in the neutrino sector, defined in Eq.(71) as $\Delta m^2_{(\nu)} \equiv (\Delta_{hm} \cdot \Delta_{ml} \cdot \Delta_{hl})_{(\nu)}$.
 Consequently, this model facilitates an estimation of the leptogenesis magnitude and its comparison with baryogenesis,
 revealing that leptogenesis is at least 71 orders of magnitude weaker than baryogenesis within this framework. \\

In the fermion mass spectrum, a degeneracy between two of the three eigenvalues emerges as $\bf C$ approaches zero.
However, the mechanism by which $\bf C$ acquires a non-trivial value remains under investigation; it is speculated to be related to the cooling of the universe during its expansion.
Furthermore, all three eigenvalues become degenerate when both $\bf B$ and $\bf C$ (i.e., $\beta$ and $\gamma$) approach zero, and when $g$ also tends to 1.
This indicates that CP symmetry violation is closely tied to the breaking of $S_N$ symmetry, but not necessarily to the presence of mass degeneracy. \\

In Section III-A, six potential correspondences are analyzed between two experimentally determined MSDs, $\Delta_a$ and $\Delta_b$, and the three theoretically defined quantities $\Delta_{hm}$, $\Delta_{ml}$, and $\Delta_{hl}$.
 Two of these correspondences are ruled out due to inconsistencies, leaving four viable cases for further study.\\

In Subsection III-B, the MSDs are analyzed across all four viable cases.
Subsequently, all twelve MSDs for the four fermion types are determined, enabling the evaluation of leptogenesis in the lepton sector and baryogenesis in the quark sector.
By substituting these results into the Jarlskog measure of CPV and incorporating the current experimental estimate of the leptonic Jarlskog invariant $J_{(l)}$,
it is found that the magnitude of leptogenesis is at least 71 orders of magnitude smaller than that of baryogenesis within this CPVSM framework. \\

In Section III-C, analytical expressions are derived to describe how the masses $m_h$, $m_m$, $m_l$, and the ratio $g' \equiv {m_m \over  m_l}$ vary as functions of the mass ratio $g \equiv {m_h \over m_m}$ across these four cases. 
Among them, two cases—Class 1—yield neutrino mass predictions restricted to narrow ranges: $1 < g < 1.01512$ for $\bf Case~1$ and $1 < g < 1.01467$ for $\bf Case~ 5$, suggesting that $m_m$ is closer in value to $m_h$.
The remaining two cases—Class 2—allow for broader ranges: $1 < g < 5.81614$ for $\bf Case~ 2$ and $1 < g < 5.90148$ for $\bf Case~4$, indicating that $m_m$ is closer in value to $m_l$. \\

As a result, all four cases predict similar values for the heaviest neutrino mass, $m_h \approx 5.01 \times 10^{-2}$ eV, 
and the lightest neutrino mass, $m_l \approx 6.09098 \times 10^{-3}$ eV, as $g$ approaches the $g$ values shown in Sunsection III-A.
For the middle neutrino mass, the model offers two possible values: 
either $m_m \approx 4.973 \times 10^{-2}$ eV if $m_m$ is closer to $m_h$, 
 or $m_m \approx 1.015 \times 10^{-2}$ eV if $m_m$ is closer to $m_l$. 
The predicted value of $m_l \approx 6.09098 \times 10^{-3}$ eV differs slightly from the value of $m_1 = 8.61 \times 10^{-3}$ eV, 
as reported in \cite{Esteban2020}, which corresponds to the square root of $\Delta_b$.  
Moreover, in each case, there exists a distinct region of interest situated between the blue and green points in Figs. 2, 4, 6, and 8.
These predictions are expected to be testable in the near future through ongoing or planned experiments.  \\
 
In summary, this article  explores potential degeneracies of the mass eigenvalues in the CPVSM and provides predictions for neutrino masses based on two experimentally given MSDs. 
In addition to the heaviest and lightest neutrinos, the mass of the middle neutrino is also estimated.
 These theoretical predictions are anticipated to be confirmed by ongoing or upcoming experiments, contributing to a deeper understanding of neutrino mass hierarchies and CP violation in the lepton sector. 
As a side-effect, the strength of leptogenesis is  also investigated as all MSDs are available, and the result shows it is negligible when compared to baryongemesis in the Standard Model.
That reveals a need of physics Beyond the Standard Model if one expects leptogenesis contribute significantly to the Baryon Asymmetry of the Universe. \\

$\bf Acknowledgement$

The author is grateful to $\bf Claude$ (Anthropic) and $\bf ChatGPT$ for assistance with manuscript revision, formatting, and language refinement.

\end{document}